\documentclass[12pt, draftclsnofoot, onecolumn]{IEEEtran}
\IEEEoverridecommandlockouts
\usepackage{cite}
\usepackage{amsmath,amssymb,amsfonts,mathabx}
\usepackage{graphicx}
\usepackage{textcomp}
\usepackage{xcolor,capt-of}
\usepackage{algorithmic,algorithm,multirow}

\DeclareMathOperator{\Tr}{\mathrm{Tr}}
\DeclareMathOperator{\Rank}{\mathrm{Rank}}
\DeclareMathOperator{\Diag}{\mathrm{Diag}}
\newcommand{\Nt}{{N_\mathrm{t}}}

\newcommand\relphantom[1]{\mathrel{\phantom{#1}}}
\allowdisplaybreaks

\newtheorem{lem}{Lemma}
\newtheorem{prop}{Proposition}
\newtheorem{thm}{Theorem}

\renewcommand{\baselinestretch}{1.47}

\begin{document}

\title{IRS-Assisted Green Communication Systems: Provable Convergence and Robust Optimization
}
\author{Xianghao~Yu,~\IEEEmembership{Member,~IEEE},
	Dongfang~Xu,~\IEEEmembership{Student Member,~IEEE},
	Derrick~Wing~Kwan~Ng,~\IEEEmembership{Senior Member,~IEEE},
	and~Robert~Schober,~\IEEEmembership{Fellow,~IEEE}
	\thanks{This work will be presented in part at the IEEE Global Communications Conference, Taipei, Taiwan,  Dec. 2020 \cite{my}.
		
		X. Yu, D. Xu, and R. Schober are with the Institute for Digital Communications, Friedrich-Alexander-University ErlangenNurnberg, 91054 Erlangen, Germany (e-mail: \{xianghao.yu, dongfang.xu, robert.schober\}@fau.de).
		
		D. W. K. Ng is with the School of Electrical Engineering and Telecommunications, University of New South Wales, Sydney,
		NSW 2052, Australia (e-mail: w.k.ng@unsw.edu.au).
	}
}

\maketitle

\begin{abstract}
Intelligent reflecting surfaces (IRSs) are regarded as a key enabler of green wireless communication, due to their capability of customizing favorable wireless propagation environments. In this paper, we investigate resource allocation for  IRS-assisted green multiuser multiple-input single-output (MISO) systems. 
To minimize the total transmit power, both the beamforming vectors at the access point (AP) and the phase shifts at multiple IRSs are jointly optimized, while taking into account the minimum required quality-of-service (QoS) of multiple users. 
First, two novel algorithms, namely a penalty-based alternating minimization (AltMin) algorithm and an inner approximation (IA) algorithm, are developed to tackle the non-convexity of the formulated optimization problem when perfect channel state information (CSI) is available. Unlike existing designs that cannot ensure convergence, the proposed penalty-based AltMin and IA algorithms are guaranteed to converge to a stationary point and a Karush-Kuhn-Tucker (KKT) solution of the design problem, respectively. 
Second, the impact of  imperfect knowledge of the CSI of the channels between the AP and the users is investigated. To this end, a non-convex robust optimization problem is formulated and the penalty-based AltMin algorithm is extended to obtain a stationary solution.
Simulation results reveal a key trade-off between the speed of convergence  and the achievable total transmit power for the two proposed algorithms.
In addition, we show that 
the proposed algorithms can significantly reduce the total transmit power at the AP compared to various baseline schemes and that 
the optimal numbers of transmit antennas and IRS reflecting elements, which minimize the total power consumption of the considered system, are finite.

\end{abstract}


\section{Introduction}
Green wireless communications has received considerable attention over the past decades and aims to reduce the power consumption of wireless networks  \cite{9113273}. In particular, various technologies for green communications have been proposed including cloud radio access networks (C-RANs) \cite{6786060}, energy harvesting \cite{6951347}, and cooperative relaying \cite{6065681}. However, these existing approaches share two common disadvantages. First, the deployment of centralized baseband unit (BBU) pools,  hardware components for energy harvesting, and active relays inevitably causes additional power consumption. Second, the performance of these wireless communication systems is still limited by the wireless channels which are treated as a ``black box" and cannot be adaptively controlled as would be desirable for green communications. Therefore,  for next-generation wireless communications, a new paradigm, which allows the customization of the wireless propagation environment, is needed to overcome these two demerits.

Recently, thanks to the  development of advanced radio frequency (RF) micro-electro-mechanical systems (MEMS), the integration of intelligent reflecting surfaces (IRSs) into wireless communication systems has been  proposed \cite{di2019smart}. In particular, employing programmable reflecting elements, IRSs are able to control the reflections of  impinging wireless signals \cite{8466374}. This unique property enables the customization of favorable wireless propagation environments, which can be exploited for further reduction of the power consumption of wireless systems. More importantly, IRSs typically require only a small amount of power for their operation as the reflecting elements are implemented by \emph{passive} hardware components, e.g., dipoles and phase shifters \cite{8990007}.
Furthermore, IRSs can be fabricated as artificial thin films attached to existing infrastructures, such as the facades of buildings, overpasses, and smart t-shirts \cite{di2019smart}, which greatly reduces the implementation cost. Indeed, IRSs are considered as a promising candidate for realizing power-efficient green wireless communications and are cost-effective devices that enable the manipulation of radio propagation environments \cite{8644519,9110912}. 
However, to fully exploit the capabilities of IRSs for reducing the power consumption of wireless systems, the IRS phase shifts have to be delicately designed and integrated with conventional communication techniques, such as the transmit beamforming at the access point (AP).

To unleash the potential of IRSs for facilitating green communications,  several works have focused on resource allocation design.
For instance, the energy efficiency of an IRS-assisted system was maximized in \cite{8741198}, where  suboptimal zero-forcing beamforming was assumed at the AP. Hence, a significant performance loss is expected as the joint design of the beamformers and reflecting elements was not considered. Besides, transmit power minimization was investigated for multiuser multiple-input single-output (MISO) systems \cite{8811733}, Internet-of-Things (IoT) applications \cite{9013643}, and simultaneous wireless information and power transfer (SWIPT) systems \cite{xujie}. Specifically, in \cite{8811733,9013643,xujie}, based on alternating minimization (AltMin) and semidefinite relaxation (SDR) methods, the total transmit power of the system was minimized while taking into account the minimum required quality-of-service (QoS) of the users.
However, the application of the SDR approach to solve a feasibility problem during AltMin does not guarantee the feasibility of the obtained solution. 
In particular, the rank-one solutions generated by the Gaussian randomization process are not guaranteed to satisfy the QoS constraints. Therefore, the monotonic convergence of  SDR-based AltMin algorithms is not ensured. 
In other words, computationally-efficient algorithms with  convergence guarantees are still an open problem for green IRS-empowered wireless communication system design.

On the other hand, as IRSs are typically implemented without  power-hungry RF chains, it is not possible to directly estimate the reflected channels by regarding the IRSs as conventional RF chain-driven transceivers. Hence, in practice, there inevitably exist non-negligible channel estimation errors when acquiring the channel state information (CSI), which have to be taken into account for the design of IRS-assisted wireless systems. To this end, robust optimization for IRS-empowered wireless systems has been investigated in \cite{9110587,xu2020resource,9133130,hong2020robust,zhao2020outage,zhang2020robust,zhou2020framework}. For example, imperfect CSI knowledge of the reflected channels between the IRS and the users was assumed in \cite{9110587,xu2020resource}. However, the two-hop reflected channels, i.e., the AP-IRS and IRS-user channels, are typically cascaded to form one effective end-to-end channel for channel estimation  \cite{hu2019two,9104260}, and hence, the CSI of the individual reflected links and the corresponding estimation errors should not be considered separately.
Hence, the imperfect CSI of the cascaded reflected channels and the direct channels between the AP and users was jointly taken into consideration in \cite{zhao2020outage,zhang2020robust,zhou2020framework}.
Specifically, an outage-constrained robust optimization problem was considered in \cite{zhao2020outage} based on an approximation of the outage probability induced by the CSI errors.
In addition,  worst-case robust optimization was studied in \cite{zhang2020robust,zhou2020framework} for different practical system settings, where, however, lower bounds on the received signal power were used to facilitate tractable robust optimization. 
Hence, an accurate and comprehensive robust optimization framework for IRS-assisted wireless systems and  corresponding provably convergent algorithms are not available in the literature, yet.

To address the aforementioned issues, this paper studies the power-efficient and robust resource allocation for IRS-assisted green multiuser MISO systems, where a multi-antenna AP serves multiple users with the help of multiple IRSs that are implemented by programmable phase shifters. 
We investigate the joint design of the beamforming vectors at the AP and the phase shifts at the IRSs for the minimization of the total transmit power, while guaranteeing a minimum required signal-to-interference-plus-noise ratio (SINR) at each user.
In the first step, we study the resource allocation for perfect CSI at the AP. Instead of applying the conventional SDR approach employed in the literature, two iterative algorithms are proposed for tackling the non-convexity of the formulated optimization problem. 
In particular, the unit modulus constrained problem is reformulated as a rank-constrained problem, which is then handled by a novel equivalent difference of convex (d.c.) functions representation. 
The first proposed algorithm, referred to as the penalty-based AltMin algorithm, leverages the penalty-based method and successive convex approximation (SCA) to address the non-convex d.c. term. 
In the second proposed algorithm, which employs inner approximation (IA), a convex program is solved in each iteration by convexifying the non-convex d.c. constraints.
Unlike existing  algorithms that cannot guarantee convergence and local optimality \cite{8811733,9013643,xujie}, the proposed penalty-based AltMin and IA algorithms are guaranteed to converge to a stationary point and a Karush-Kuhn-Tucker (KKT) solution of the considered non-convex power minimization problem, respectively. 
Moreover, in the  second step,  the imperfect CSI of both the reflected and direct channels is taken into account, and the robust resource allocation design is formulated as another non-convex optimization problem. 
For this case, the penalty-based AltMin algorithm developed for perfect CSI is extended to efficiently account for the imperfect CSI and to attain a stationary point of the formulated robust non-convex optimization problem. 
Simulation results reveal that IRSs are an effective enabler of green wireless communications in terms of reducing the total transmit power at the AP. It is shown that the new algorithms proposed for perfect CSI outperform the state-of-the-art SDR-based AltMin algorithm, and for imperfect CSI, the robustness of the respective proposed AltMin algorithm is also confirmed. In addition, for the perfect CSI case, the proposed penalty-based algorithm enjoys a faster convergence at the expense of a higher total transmit power than the proposed IA algorithm, which reveals a critical trade-off between  performance and algorithm convergence rate for green IRS-assisted wireless systems.

\emph{Notations:} In this paper, $\jmath=\sqrt{-1}$ denotes the imaginary unit of a complex number.
Vectors and matrices are denoted by boldface lower-case and  capital letters, respectively.
The set of nonnegative integers is denoted as $\mathbb{N}=\{0,1,\cdots\}$.
$\mathbb{C}^{m\times n}$ stands for the set of all $m\times n$ complex-valued matrices; 
$\mathbb{H}^{m}$ represents the set of all $m\times m$ Hermitian matrices; 
$\mathbf{1}_m$ denotes the $m\times1$ all-ones vector; $\mathbf{I}_m$ is the $m$-dimensional identity matrix.
$\mathbf{X}^*$, $\mathbf{X}^T$, and $\mathbf{X}^H$ stand for the conjugate, transpose, and conjugate transpose of matrix $\mathbf{X}$, respectively.
The $i$-th element of vector $\mathbf{x}$ is denoted as $x_i$. 
The $\ell_2$-norm of vector $\mathbf{x}$ is denoted as $\left\Vert\mathbf{x}\right\Vert_2$. The spectral norm, nuclear norm, and Frobenius norm  of matrix $\mathbf{X}$ are represented as $\left\Vert\mathbf{X}\right\Vert_2$, $\left\Vert\mathbf{X}\right\Vert_*$, and $\left\Vert\mathbf{X}\right\Vert_F$, respectively.
$\mathrm{diag}(\mathbf{x})$ represents a diagonal matrix whose main diagonal elements are extracted from vector $\mathbf{x}$,
and $\mathrm{blkdiag}\left(\mathbf{X}_1,\cdots,\mathbf{X}_n\right)$ denotes a block diagonal
matrix whose diagonal components are $\mathbf{X}_1,\cdots,\mathbf{X}_n$.
$\mathrm{Diag}(\mathbf{X})$ denotes  a vector whose elements are extracted from the main diagonal elements of matrix $\mathbf{X}$.
The largest eigenvalue of matrix $\mathbf{X}$ and its associated eigenvector are denoted by  ${\lambda_{\max}}\left(\mathbf{X}\right)$ and $\boldsymbol{\lambda}_{\max}(\mathbf{X})$, respectively.
$\otimes$ stands for the  Kronecker product between two matrices;
$\mathrm{det}(\mathbf{X})$, $\Rank(\mathbf{X})$, and $\Tr(\mathbf{X})$  denote the determinant, rank,  and trace of matrix $\mathbf{X}$;
$\mathbf{X}\succeq\mathbf{0}$ indicates that $\mathbf{X}$ is a positive semidefinite  matrix.
For a real-valued continuous function $f(\mathbf{X})$, $\nabla_\mathbf{X}f$ denotes the gradient of $f$ with respect to matrix $\mathbf{X}$. 
$\mathbb{E}[\cdot]$  and $\Re(\cdot)$  stand for statistical expectation and the real part of a complex number, respectively;
$\mathrm{vec}\left({\mathbf{X}}\right)$ represents the vectorization of matrix $\mathbf{X}$; 
$\mathbf{X}^\mathrm{opt}$ denotes the  optimal value of an optimization variable $\mathbf{X}$;
$\mathrm{unt}(\mathbf{x})$ forms a vector whose elements are given by $\frac{x_1}{|x_i|},\cdots,\frac{x_n}{|x_n|}$.

\section{System Model and Existing Approach}
In this section, we first present the considered IRS-assisted multiuser MISO system and the resource allocation problem when perfect CSI is available at the AP. Then, we discuss the existing approach for handling the problem and its main limitations. 
\subsection{IRS-Assisted System Model and Problem Formulation}
\begin{figure}[t]
	\centering\includegraphics[height=5cm]{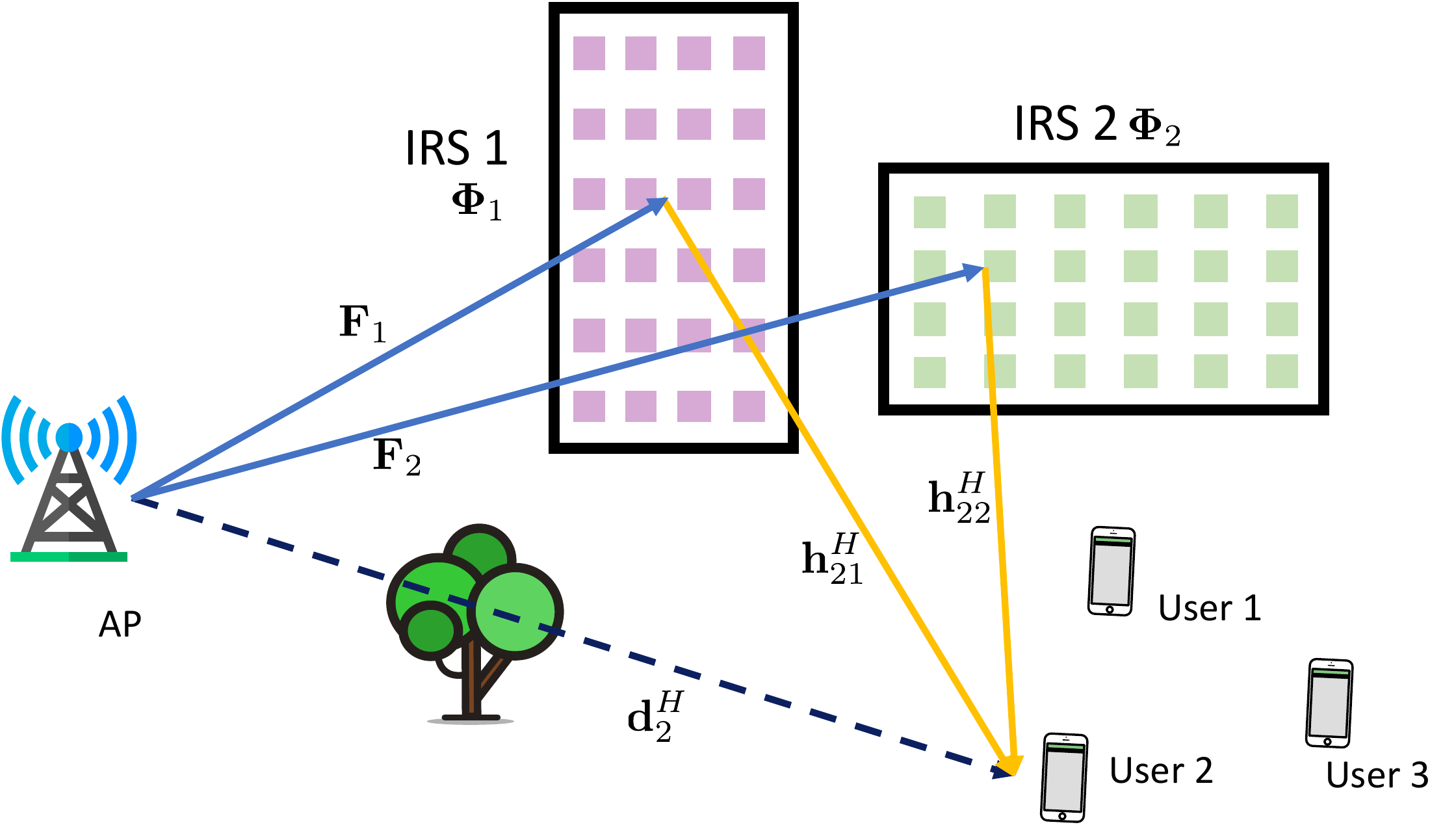}
	\caption{An IRS-assisted multiuser MISO system comprising $K=3$ users and $L = 2$ IRSs.
		For  ease of illustration, only the parameters of the channels
		of user 2 are shown.}
	\label{model}
\end{figure}

We consider the downlink transmission in an IRS-assisted multiuser MISO wireless communication system, which consists of an $\Nt$-antenna AP, $K$ single-antenna users, and $L$ IRSs, as shown in Fig. \ref{model}.
The baseband signal received at user $k$ is given by\footnote{For multiple IRSs,  the delays between the propagation paths reflected by different IRSs are typically much shorter than the symbol duration. For example, in a small cell network with $200$ m cell radius, the maximum delay is $1.3$ $\mu$s while the symbol duration in the Long-Term Evolution (LTE)  standard is $70$ $\mu$s \cite{arunabha2010fundamentals}. Thus, intersymbol interference is not considered in \eqref{signal}.}
\begin{equation}\label{signal}
y_k=\left(\sum_{l\in\mathcal{L}}\mathbf{h}_{kl}^H\mathbf{\Phi}_l\mathbf{F}_l+\mathbf{d}_k^H\right)\sum_{j\in\mathcal{K}}\mathbf{w}_js_j+n_k=\left(\mathbf{h}_k^H\mathbf{\Phi}\mathbf{F}+\mathbf{d}_k^H\right)\sum_{j\in\mathcal{K}}\mathbf{w}_js_j+n_k,\quad\forall k\in\mathcal{K},
\end{equation}
where $\mathcal{L}=\{1,\cdots,L\}$, $\mathcal{K}=\{1,\cdots,K\}$, $\mathbf{h}_k^H=\left[\mathbf{h}_{k1}^H,\cdots,\mathbf{h}_{kL}^H\right]$,
$\mathbf{\Phi}=\mathrm{blkdiag}\left(\mathbf{\Phi}_1,\cdots,\mathbf{\Phi}_l\right)$, and
$\mathbf{F}^H=\left[\mathbf{F}_1^H,\cdots,\mathbf{F}_L^H\right]$.
The channel vectors from the AP and  IRS $l$  to user $k$ are represented by $\mathbf{d}_k\in\mathbb{C}^{\Nt\times1}$ and $\mathbf{h}_{kl}\in\mathbb{C}^{M_l\times1}$, respectively. The channel matrix from the AP to IRS $l$ is denoted by $\mathbf{F}_l\in\mathbb{C}^{M_l\times\Nt}$. We assume that  IRS $l$ is implemented by $M_l$ programmable phase shifters and the phase shift matrix at  IRS $l$ is given by $\mathbf{\Phi}_l=\mathrm{diag}\left(e^{\jmath\theta_{l1}},\cdots,e^{\jmath\theta_{l{M_l}}}\right)$, where $\theta_{lm}\in[0,2\pi]$, $\forall m\in\{1,\cdots,M_l\}$, represents the phase shift of the $m$-th reflecting element at IRS $l$. 
The information-carrying signal transmitted from the AP to user $j$ is denoted by $s_j$, where $\mathbb{E}\left[|s_j|^2\right]=1$, $\forall j\in\mathcal{K}$, without loss of generality. The beamforming vector for user $j$ is denoted by $\mathbf{w}_j$. Variable $n_k$ represents the additive white Gaussian noise at user $k$ with zero mean and variance $\sigma_k^2$.
For notational convenience, in the rest of this paper, we drop  index $l$ in the elements of the  phase shift matrices of all IRSs and use $\mathbf{\Phi}=\mathrm{diag}\left(\boldsymbol{\phi}\right)$, where $\boldsymbol{\phi}\triangleq\left[e^{\jmath \theta_{1}},e^{\jmath \theta_{2}},\cdots,e^{\jmath \theta_{M}}\right]^T$ and $M=\sum_{l\in\mathcal{L}}M_l$.
Therefore, the received SINR at user $k$ is given by
\begin{equation}\label{sinr}
\mathrm{SINR}_k=\frac{\left|\left(\mathbf{h}_k^H\mathbf{\Phi F}+\mathbf{d}_k^H\right)\mathbf{w}_k\right|^2}
{\sum_{j\in\mathcal{K}\backslash\{k\}}\left|\left(\mathbf{h}_k^H\mathbf{\Phi F}+\mathbf{d}_k^H\right)\mathbf{w}_j\right|^2+\sigma_k^2}.
\end{equation}

Our goal in this paper is to minimize the total transmit power at the AP while ensuring a minimum required QoS of the users. In Sections II and III, we first assume that the full CSI of  the considered IRS-assisted system is available for resource allocation design, which leads to a theoretical performance upper bound for the
considered system. Then, the impact of  CSI uncertainty will be taken into account in  Section IV.
Now, with full CSI, the proposed power-efficient design of the beamformers at the AP and the reflecting elements at the IRSs is obtained by solving the following optimization problem:
\begin{equation}
\begin{aligned}
&\underset{\mathbf{w}_k,\mathbf{\Phi}}{\mathrm{minimize}} && f\left(\mathbf{w}_k\right)=\sum_{k\in\mathcal{K}}\left\Vert\mathbf{w}_k\right\Vert_2^2\\
&\mathrm{subject\thinspace to}&&\mbox{C1:}\,\mathrm{SINR}_k\ge\gamma_k,\quad\forall k,\quad\mbox{C2:}\,\mathbf{\Phi}=\mathrm{diag}\left(
e^{\jmath\theta_1},e^{\jmath\theta_2},\cdots,e^{\jmath\theta_M}\right),
\end{aligned}\label{problem}
\end{equation}
where $\gamma_k$ is the predefined minimum required SINR of user $k$.

\emph{Remark 1:} There are two main challenges in solving problem \eqref{problem}. First, each IRS reflecting element in $\mathbf{\Phi}$ has a unit modulus, i.e., $\left|e^{\jmath\theta_m}\right|=1$, $\forall m\in\{1,\cdots,M\}$, which intrinsically is a highly non-convex constraint. Second, the optimization variables $\mathbf{w}_k$ and $\mathbf{\Phi}$ are coupled in  QoS constraint \mbox{C1}. These two aspects complicate problem \eqref{problem} as it is not jointly convex with respect to the optimization variables, and hence, in general difficult to  solve optimally.

\subsection{Existing Approach}\label{IIB}
To tackle the difficulties in solving problem \eqref{problem}, SDR-based AltMin algorithms have been widely adopted in the literature \cite{8811733,9013643,xujie} to obtain a suboptimal solution. In particular,  the optimization of $\mathbf{w}_k$ and $\mathbf{\Phi}$ is decoupled and performed alternately by capitalizing on AltMin. For a fixed $\mathbf{\Phi}$, the beamformers $\mathbf{w}_k$ are optimized based on
\begin{equation}\label{p1}
\begin{aligned}
&\underset{\mathbf{w}_k}{\mathrm{minimize}} && f\left(\mathbf{w}_k\right)=\sum_{k\in\mathcal{K}}\left\Vert\mathbf{w}_k\right\Vert_2^2\\
&\mathrm{subject\thinspace to}&&\mbox{C1:}\,\mathrm{SINR}_k\ge\gamma_k,\quad\forall k,
\end{aligned}
\end{equation}
which is identical to the classic beamforming problem in wireless systems without IRSs and therefore can be optimally solved via second-order cone programming  \cite{8811733}. 
On the other hand, the phase shift matrix $\mathbf{\Phi}$ can be optimized by solving the following feasibility check problem:
\begin{equation}\label{eq4}
\begin{aligned}
&\underset{\mathbf{\Phi}}{\mathrm{minimize}} && 1\\
&\mathrm{subject\thinspace to}&&\mbox{C1:}\,\mathrm{SINR}_k\ge\gamma_k,\quad\forall k,\quad\mbox{C2:}\,\mathbf{\Phi}=\mathrm{diag}\left(
e^{\jmath\theta_1},e^{\jmath\theta_2},\cdots,e^{\jmath\theta_M}\right).
\end{aligned}
\end{equation}

However, even if problem \eqref{eq4} can be solved optimally, the monotonicity of the objective value during the AltMin iteration cannot be guaranteed. Specifically, in the $t$-th iteration, the optimal beamformers, $\mathbf{w}_k$, conditioned on a given  $\mathbf{\Phi}^{(t)}$ are obtained by solving problem \eqref{p1}, and are denoted by $\mathbf{w}_k^{(t)}|\mathbf{\Phi}^{(t)}$. 
Unfortunately, we have
\begin{equation}\label{eq8}
f\left(\mathbf{w}_k^{(t+1)}|\mathbf{\Phi}^{(t+1)}\right){\not\leq}f\left(\mathbf{w}_k^{(t)}|\mathbf{\Phi}^{(t)}\right).
\end{equation}
This is because for problem $\eqref{p1}$, $\mathbf{w}_k^{(t)}|\mathbf{\Phi}^{(t)}$ and $\mathbf{w}_k^{(t+1)}|\mathbf{\Phi}^{(t+1)}$  are two optimal solutions conditioned on two  different sets of parameters $\mathbf{\Phi}^{(t)}$ and $\mathbf{\Phi}^{(t+1)}$, respectively. 
When the parameters of problem $\eqref{p1}$ are updated from $\mathbf{\Phi}^{(t)}$ to $\mathbf{\Phi}^{(t+1)}$, the feasible set  changes. However, the relation between the two feasible sets are difficult to quantify as  $\mathbf{\Phi}^{(t)}$ and $\mathbf{\Phi}^{(t+1)}$ are merely feasible solutions of problem \eqref{eq4}, which are probably not unique.
Hence, there is no guarantee that the optimal objective value  achieved by $\mathbf{w}_k^{(t+1)}$ is smaller than that for $\mathbf{w}_k^{(t)}$, as stated in \eqref{eq8}, which forms the first limitation of the existing SDR-based AltMin algorithm.

Moreover, problem \eqref{eq4} cannot be solved optimally. In particular, the main challenge lies in the non-convex unit modulus constraints in $\mathbf{\Phi}$, which are typically handled by the SDR approach in the literature \cite{8811733,9013643,xujie}.
According to \cite[eq. (44)]{8811733}, problem \eqref{eq4} can be reformulated as
\begin{equation}\label{eq5}
\begin{aligned}
&\underset{\mathbf{V}\in\mathbb{H}^{M+1}\succeq\mathbf{0}}{\mathrm{minimize}} && 1\\
&\mathrm{subject\thinspace to}&&\widecheck{\mbox{C1}}\mbox{:}\,\Tr\left(\mathbf{R}_k\mathbf{V}\right)\le\xi_k,\quad\forall k,
\quad\widehat{\mbox{C2}}\mbox{:}\,\Diag\left(\mathbf{V}\right)=\mathbf{1}_{M+1},\quad \mbox{C3:}\,\Rank\left(\mathbf{V}\right)=1,
\end{aligned}
\end{equation}
where $\xi_k=\left|\mathbf{d}_k^H\mathbf{w}_k\right|^2-\gamma_k\left(\sigma_k^2+\sum_{j\in\mathcal{K}\backslash\{k\}}\left|\mathbf{d}_k^H\mathbf{w}_j\right|^2\right)$, $\mathbf{v}=\left[\boldsymbol{\phi}^T,x\right]^H$, $x$ with $|x|^2=1$ is an auxiliary optimization variable,  and $\mathbf{V}=\mathbf{vv}^H$. In addition, $\mathbf{R}_k=-\mathbf{T}_{k,k}+\gamma_k\sum_{j\in\mathcal{K}\backslash\{k\}}\mathbf{T}_{k,j}$, where $\mathbf{T}_{k,j}$ is given by 
\begin{equation}
\mathbf{T}_{k,j}=\begin{bmatrix}
\mathrm{diag}\left(\mathbf{h}_k^H\right)\mathbf{F}\mathbf{w}_j\mathbf{w}_j^H\mathbf{F}^H\mathrm{diag}\left(\mathbf{h}_k\right)&\mathrm{diag}\left(\mathbf{h}_k^H\right)\mathbf{F}\mathbf{w}_j\mathbf{w}_j^H\mathbf{d}_k\\
\mathbf{d}_k^H\mathbf{w}_j\mathbf{w}_j^H\mathbf{F}^H\mathrm{diag}\left(\mathbf{h}_k\right)&0
\end{bmatrix}.
\end{equation}
By adopting SDR to handle problem \eqref{eq5}, the non-convex rank-one constraint \mbox{C3} is dropped. Then, the relaxed problem  becomes a semidefinite programming (SDP), which can be solved by standard convex program solvers such as CVX \cite{grant2008cvx}.  

\begin{figure}[t]
	\centering
	\begin{minipage}[t]{0.485\linewidth} \hspace{-1em}
		\centering\includegraphics[height=6cm]{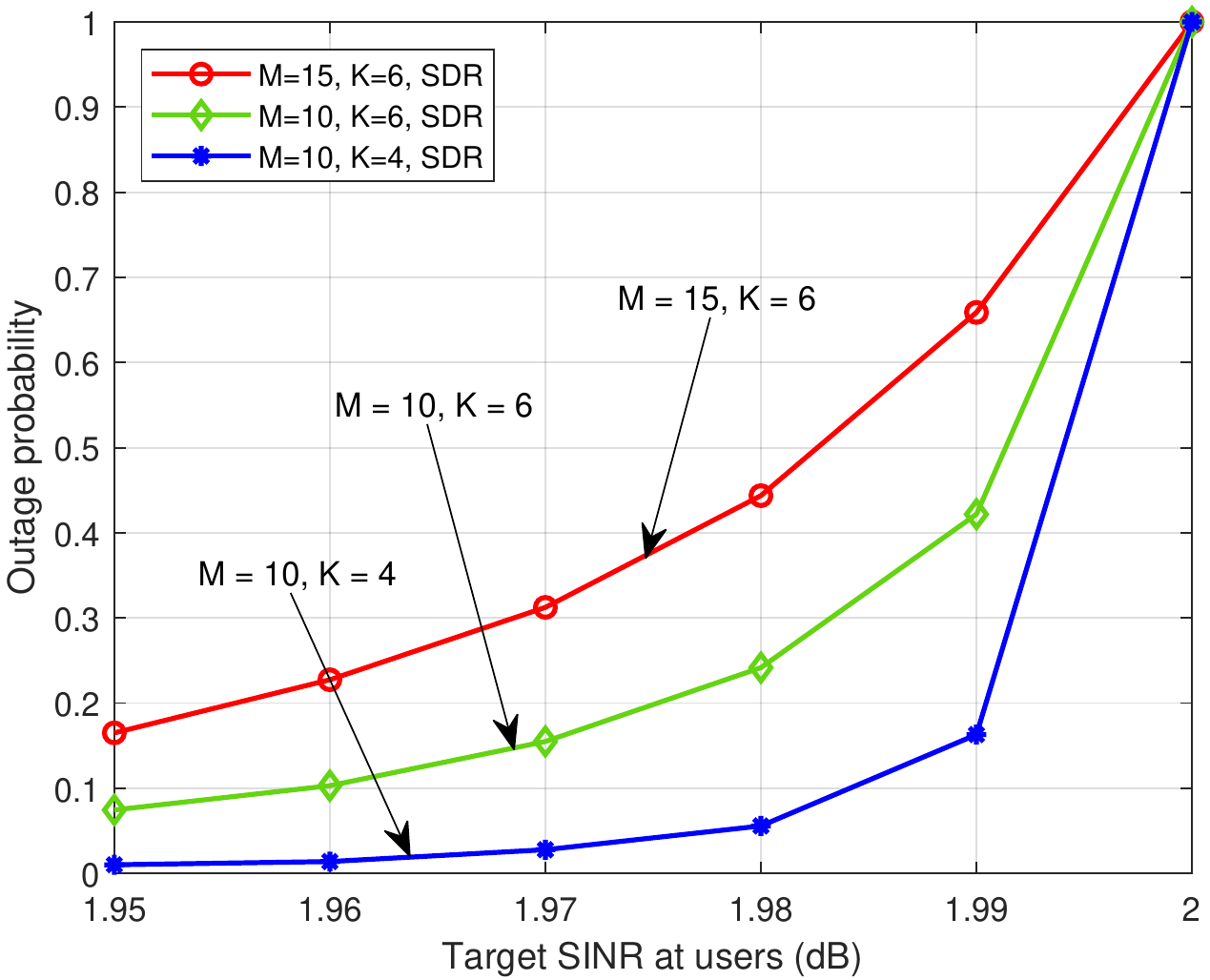}
		\caption{Outage probability of QoS constraint \mbox{C1} when solving problem \eqref{eq5} via SDR. The parameters are set as $\Nt=6$, $\gamma_k=2$ dB,
			 $\tilde{\gamma}_{k}=\tilde{\gamma}$, 
			and $\sigma_k^2=-90$ dBm.}
		\label{fig0}
	\end{minipage}\quad
	\begin{minipage}[t]{0.485\linewidth} \hspace{-1em}
		\centering\includegraphics[height=6cm]{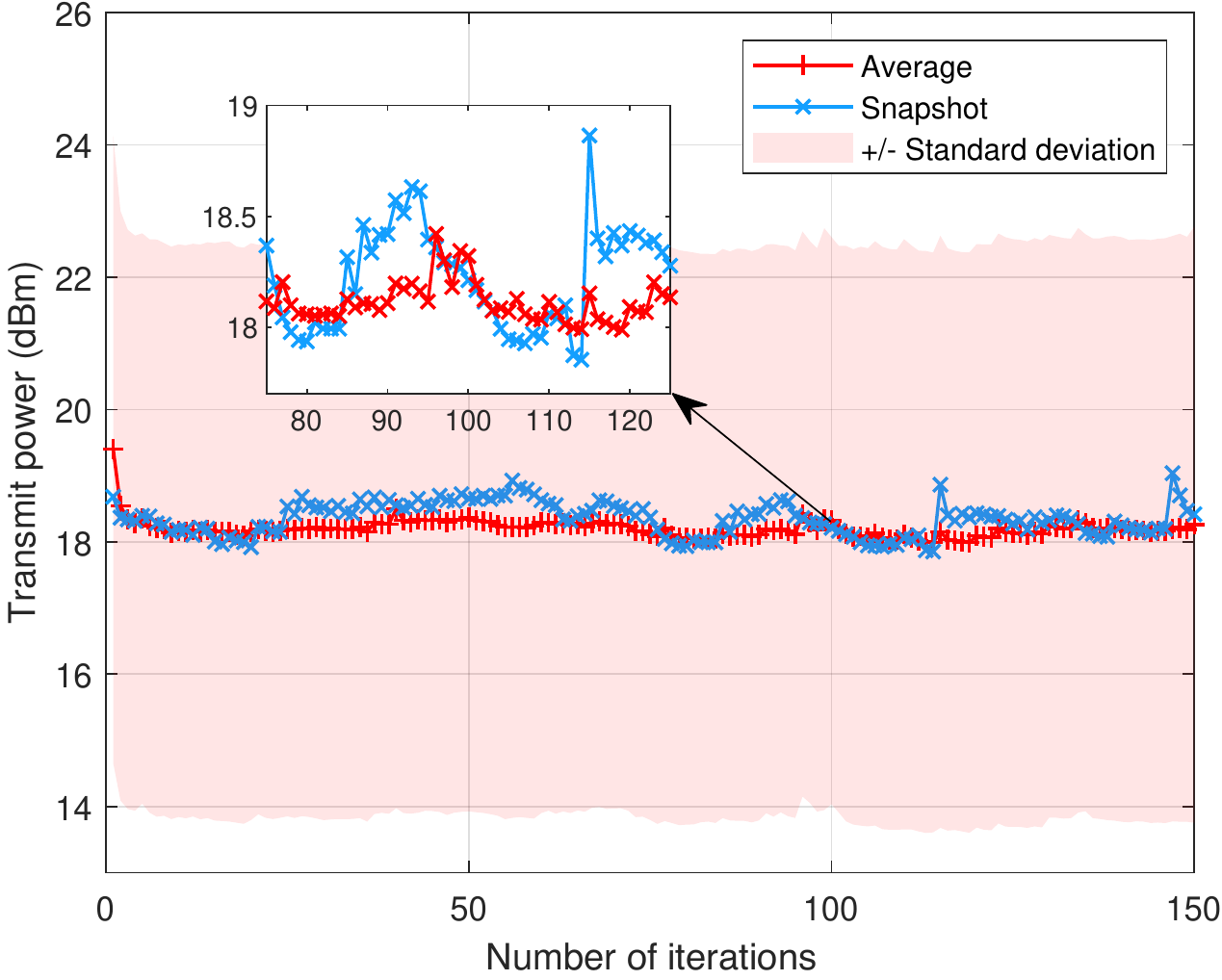}
		\caption{Convergence of the SDR-AltMin algorithm for $\Nt=M=10$, $K=3$, $\gamma_k=2$ dB, and $\sigma_k^2=-90$ dBm. 
			The shaded area represents the +/- standard deviation of the convergence curves over all realizations.
		}
		\label{fig1}
	\end{minipage}
\end{figure}

Unfortunately, there is no guarantee that the obtained optimal solution $\mathbf{V}^\mathrm{opt}$ is a rank-one matrix. Hence, Gaussian randomization  can be utilized  to satisfy constraint \mbox{C3}. In particular, a Gaussian random vector $\tilde{\mathbf{v}}$ is generated with zero mean and covariance matrix $\mathbf{V}^\mathrm{opt}$. 
To satisfy  constraint $\widehat{\mbox{C2}}$ in \eqref{eq5}, the randomized vector $\tilde{\mathbf{v}}$ is thus normalized to a unit modulus vector  $\mathbf{v}=\mathrm{unt}\left(\tilde{\mathbf{v}}\right)$. 
Nevertheless, it is not guaranteed that this normalized vector $\mathbf{v}$ satisfies the QoS constraint $\widecheck{\mbox{C1}}$ in \eqref{eq5}, and, equivalently, constraint \mbox{C1} in \eqref{eq4}.
To illustrate the drawback of this method,  we evaluate the feasibility  of the normalized random vector $\mathbf{v}$ via the outage probability, defined as the probability that the received SINRs at the users are lower than a predefined target SINR. In Fig. \ref{fig0}, we set the minimum required SINR of the users to $\gamma_{k}=2$ dB. 
As can be observed,  the outage probability approaches one when the target SINR is $2$ dB, which indicates that the normalized Gaussian randomized vector $\mathbf{v}$ can almost never guarantee the required QoS. The failure to guarantee a feasible solution is the second limitation of the existing SDR-based AltMin algorithm. 
Inspired by Fig. \ref{fig0}, one can  manually increase the minimum required SINR $\gamma_k$ by a small value when solving problem \eqref{eq5}, so that the randomized solution $\mathbf{v}$ fulfills the original QoS constraint \mbox{C1} with  high  probability. However, this heuristic approach modifies the feasible set of the original optimization problem and hence the optimality of the resulting solution and the monotonicity of convergence cannot be guaranteed.

Considering the two above-mentioned limitations, both the convergence and the performance of the state-of-the-art SDR-based AltMin algorithm depend on ``luck'' due to the application of Gaussian randomization in solving the feasibility check problem \eqref{eq4}. As shown in Fig. \ref{fig1}, the objective function fluctuates significantly during AltMin.
This motivates us to develop two novel algorithms for solving problem \eqref{problem} in the next section.

\section{Algorithm Design for IRS-Assisted MISO Systems With Perfect CSI}
In this section, we propose two novel algorithms for resource allocation in IRS-empowered multiuser MISO systems. Unlike the existing SDR-based AltMin algorithm, convergence guarantees can be given  for these two proposed algorithms and local optimality is guaranteed for the latter.
\subsection{Problem Reformulation}
In this section, we reformulate problem \eqref{problem} as follows.
The numerator of the SINR in \eqref{sinr} is rewritten as
\begin{equation}
\begin{split}
\left|\left(\mathbf{h}_k^H\mathbf{\Phi F}+\mathbf{d}_k^H\right)\mathbf{w}_k\right|^2&=
2\Re\left(\boldsymbol{\phi}^T\mathbf{E}_k\mathbf{W}_k\mathbf{d}_k\right)+\boldsymbol{\phi}^T\mathbf{E}_k\mathbf{W}_k\mathbf{E}_k^H\boldsymbol{\phi}^*+\mathbf{d}^H_k\mathbf{W}_k\mathbf{d}_k\\
&=\mathbf{v}^H\mathbf{H}_k^H\mathbf{W}_k\mathbf{H}_k\mathbf{v}=\Tr\left(\mathbf{V}\mathbf{H}_k^H\mathbf{W}_k\mathbf{H}_k\right),
\end{split}
\end{equation}
where $\mathbf{E}_k=\mathrm{diag}\left(\mathbf{h}^H_k\right)\mathbf{F}$, $\mathbf{W}_k=\mathbf{w}_k\mathbf{w}_k^H$, and $\mathbf{H}_k=\begin{bmatrix}
\mathbf{E}_k^H&\mathbf{d}_k
\end{bmatrix}$ is an effective channel matrix including both the reflected channels and direct channel from the AP to user $k$. The denominator can be rewritten in a similar manner and problem \eqref{problem} is equivalently reformulated as
\begin{equation}\label{refor}
\begin{aligned}
&\underset{\mathbf{W}_k\in\mathbb{H}^{\Nt},\mathbf{V}\in\mathbb{H}^{M+1}\succeq\mathbf{0}}{\mathrm{minimize}} && f\left(\mathbf{W}_k\right)=\sum_{k\in\mathcal{K}}\Tr\left(\mathbf{W}_k\right)\\
&\quad\,\,\,\,\mathrm{subject\thinspace to}&&\widehat{\mbox{C1}}\mbox{:}\,\gamma_k\sigma_k^2+\Tr\left(\mathbf{V}\mathbf{H}_k^H\tilde{\mathbf{W}}_k\mathbf{H}_k\right)\le0,\quad\forall k,
\\
&&&\widehat{\mbox{C2}}, \mbox{C3},\quad\mbox{C4:}\,\Rank\left(\mathbf{W}_k\right)\le1,\quad\forall k,
\end{aligned}
\end{equation}
where $\tilde{\mathbf{W}}_k=\gamma_k\sum_{j\in\mathcal{K}\backslash \{k\}}\mathbf{W}_j-\mathbf{W}_k$. The constraints $\mathbf{W}_k\in\mathbb{H}^{\Nt}\succeq\mathbf{0}$ and \mbox{C4} are imposed to guarantee that $\mathbf{W}_k=\mathbf{w}_k\mathbf{w}_k^H$ still holds after optimizing $\mathbf{W}_k$. 

As we analyzed in Section \ref{IIB}, the  state-of-the-art SDR AltMin algorithm has two main limitations. First, the feasibility check problem \eqref{eq4} prevents us from guaranteeing the monotonicity of the objective value during AltMin, as shown in \eqref{eq8}. Second, the SDR approach employed to tackle the unit modulus constraints in problem \eqref{eq4} may fail to generate a feasible solution in terms of the QoS constraint \mbox{C1}. In this section, we  propose two new algorithms with convergence guarantees for tackling the reformulated problem \eqref{refor}.

\subsection{Proposed Penalty-Based AltMin Algorithm}\label{IIIB}
For the first proposed algorithm, we follow the design principle of AltMin.
However, unlike the existing SDR-AltMin approach \cite{8811733,9013643,xujie}, we avoid the feasibility check problem \eqref{eq4} in the AltMin procedure to get rid of the problem illustrated in \eqref{eq8}. First, we define $\bar{\mathbf{W}}_k=\frac{1}{{P}}\mathbf{W}_k$, $\forall k$,
where $P=\sum_{k\in\mathcal{K}}\Tr\left(\mathbf{W}_k\right)$, i.e., $\sum_{k\in\mathcal{K}}\Tr\left(\bar{\mathbf{W}}_k\right)=1$. Now, problem \eqref{refor} can be equivalently written as
\begin{equation}\label{p30}
\begin{aligned}
&\underset{P>0,\bar{\mathbf{W}}_k\in\mathbb{H}^{\Nt},\mathbf{V}\in\mathbb{H}^{M+1}\succeq\mathbf{0}}{\mathrm{minimize}} && P\\
&\quad\quad\,\,\,\mathrm{subject\thinspace to}
&&\widehat{\mbox{C2}}, \mbox{C3}, \quad\sum_{k\in\mathcal{K}}\Tr\left(\bar{\mathbf{W}}_k\right)=1,\quad\Rank\left(\bar{\mathbf{W}}_k\right)=1,\quad\forall k,\\
&&&\gamma_k\sigma_k^2+P\Tr\left(\mathbf{V}\mathbf{H}_k^H\tilde{\bar{\mathbf{W}}}_k\mathbf{H}_k\right)\le0,\quad\forall k,
\end{aligned}
\end{equation}
where $\tilde{\bar{\mathbf{W}}}_k=\gamma_k\sum_{j\in\mathcal{K}\backslash \{k\}}\bar{\mathbf{W}}_j-\bar{\mathbf{W}}_k$. Then, we apply AltMin to handle problem \eqref{p30}. Specifically, the optimization of $P$ and $\bar{\mathbf{W}}_k$ for a fixed $\mathbf{V}$ can be formulated as
\begin{equation}\label{pwk}
\begin{aligned}
&\underset{P>0,\bar{\mathbf{W}}_k\in\mathbb{H}^{\Nt}\succeq\mathbf{0}}{\mathrm{minimize}} && P\\
&\,\,\,\,\,\mathrm{subject\thinspace to}
&& \sum_{k\in\mathcal{K}}\Tr\left(\bar{\mathbf{W}}_k\right)=1,\quad\Rank\left(\bar{\mathbf{W}}_k\right)\le1,\quad\forall k,\\
&&&\gamma_k\sigma_k^2+P\Tr\left(\mathbf{V}\mathbf{H}_k^H\tilde{\bar{\mathbf{W}}}_k\mathbf{H}_k\right)\le0,\quad\forall k.
\end{aligned}
\end{equation}
By leveraging the definition $\mathbf{W}_k=P\bar{\mathbf{W}}_k$, problem \eqref{pwk} is equivalent to 
\begin{equation}\label{p33}
\underset{\mathbf{W}_k\in\mathbb{H}^{\Nt}\succeq\mathbf{0}}{\mathrm{minimize}} \,\,\,\, \sum_{k\in\mathcal{K}}\Tr\left(\mathbf{W}_k\right)\quad\quad\mathrm{subject\thinspace to}\,\,\,\,\widehat{\mbox{C1}}, \mbox{C4}.
\end{equation}
Note that problem \eqref{p33} is equivalent to problem \eqref{p1}, which is the classical beamforming problem in wireless systems without IRSs and can be solved optimally.

\emph{Remark 2:} The optimization of beamforming matrices $\mathbf{W}_k$ in problem \eqref{p33} can be directly obtained by applying AltMin to problem \eqref{refor} without the help of the normalization step $\bar{\mathbf{W}}_k=\frac{1}{{P}}\mathbf{W}_k$. However, this normalization step does help avoid the feasibility check problem when optimizing the IRS reflecting elements in $\mathbf{V}$ for given beamforming matrices, as shown in the following.

When $\bar{\mathbf{W}}_k$ is given in \eqref{p30}, the optimization of $P$ and $\mathbf{V}$ yields
\begin{equation}\label{pv}
\begin{aligned}
&\underset{P>0,\mathbf{V}\in\mathbb{H}^{M+1}\succeq\mathbf{0}}{\mathrm{minimize}} && P\\
&\quad\mathrm{subject\thinspace to}
&&\widehat{\mbox{C2}}, \mbox{C3},\quad\gamma_k\sigma_k^2+P\Tr\left(\mathbf{V}\mathbf{H}_k^H\tilde{\bar{\mathbf{W}}}_k\mathbf{H}_k\right)\le0,\quad\forall k.
\end{aligned}
\end{equation}
By defining $\bar{\mathbf{V}}=P\mathbf{V}$, problem \eqref{pv} can be further rewritten as
\begin{equation}\label{pv1}
\begin{aligned}
&\underset{P>0,\bar{\mathbf{V}}\in\mathbb{H}^{M+1}\succeq\mathbf{0}}{\mathrm{minimize}} && P\\
&\,\,\,\,\,\mathrm{subject\thinspace to}
&&\overline{\widehat{\mbox{C2}}}\mbox{:}\,\mathrm{Diag}\left(\bar{\mathbf{V}}\right)=P\mathbf{1}_{M+1}, \quad\overline{\mbox{C3}}\mbox{:}\,\Rank\left(\bar{\mathbf{V}}\right)=1,\\
&&&\gamma_k\sigma_k^2+\Tr\left(\bar{\mathbf{V}}\mathbf{H}_k^H\tilde{\bar{\mathbf{W}}}_k\mathbf{H}_k\right)\le0,\quad\forall k.
\end{aligned}
\end{equation}
As can be observed, thanks to the proposed normalization, i.e., $\bar{\mathbf{W}}_k=\frac{1}{{P}}\mathbf{W}_k$, optimization variable $P$ appears in the objective functions of both problems \eqref{pwk} and \eqref{pv1}, 
which eliminates the feasibility check problem in \eqref{eq4} and  its  demerit illustrated in \eqref{eq8}. In fact, it guarantees the monotonicity of the objective function, i.e., $P$, when the  AltMin procedure is applied to problem \eqref{p30}.

Now, the only non-convexity in problem \eqref{pv1} lies in the rank-one constraint $\overline{\mbox{C3}}$ of $\bar{\mathbf{V}}$.
Recall that the rank-one constraint was typically tackled by SDR without  guarantee of recovering a feasible solution, as illustrated in Fig. \ref{fig0}. Instead, we employ a novel representation of the rank-one constraint shown in the following lemma.
\begin{lem}
	The rank-one constraint $\overline{\mbox{C3}}$  is equivalent to constraint $\overline{\overline{\mbox{C3}}}$, which is given by
\end{lem}
\begin{equation}\label{force}
\overline{\overline{\mbox{C3}}}\mbox{:}\,\left\Vert\bar{\mathbf{V}}\right\Vert_*-\left\Vert\bar{\mathbf{V}}\right\Vert_2\le0.
\end{equation}
\begin{IEEEproof}
	For any $\mathbf{X}\in\mathbb{H}^{m}$, the inequality $\left\Vert\mathbf{X}\right\Vert_*=\sum_i{\sigma_i}\ge\left\Vert\mathbf{X}\right\Vert_2=\underset{i}{\max}\{\sigma_i\}$ holds, where $\sigma_i$ is the $i$-th singular value of $\mathbf{X}$. Equality holds if and only if $\mathbf{X}$ has unit rank. 
\end{IEEEproof}
Then, we tackle the non-convex rank-one constraint $\overline{\mbox{C3}}$ by resorting to
a penalty-based method as in \cite[Ch. 17]{nocedal2006numerical,9133130}. Specifically, by adding a penalized version of constraint $\overline{\overline{\mbox{C3}}}$ to the objective function, problem  \eqref{pv1} is transformed to
\begin{equation}\label{pv2}
\begin{aligned}
&\underset{P>0,\bar{\mathbf{V}}\in\mathbb{H}^{M+1}\succeq\mathbf{0}}{\mathrm{minimize}} && P+\frac{1}{\mu}\left(\left\Vert\bar{\mathbf{V}}\right\Vert_*-\left\Vert\bar{\mathbf{V}}\right\Vert_2\right)\\
&\,\,\,\,\,\mathrm{subject\thinspace to}
&&\overline{\widehat{\mbox{C2}}}\mbox{:}\,\mathrm{Diag}\left(\bar{\mathbf{V}}\right)=P\mathbf{1}_{M+1},\quad\gamma_k\sigma_k^2+\Tr\left(\bar{\mathbf{V}}\mathbf{H}_k^H\tilde{\bar{\mathbf{W}}}_k\mathbf{H}_k\right)\le0,\quad\forall k,
\end{aligned}
\end{equation}
where $\mu>0$ is a penalty factor penalizing the violation of constraint $\overline{\overline{\mbox{C3}}}$. The following proposition demonstrates the equivalence of problems \eqref{pv1} and \eqref{pv2}.
\begin{prop}
	Let $\bar{\mathbf{V}}_s$ be the optimal solution of problem \eqref{pv2} with penalty factor $\mu_s$. When $\mu_s$ is sufficiently small, i.e., $\mu_s\to0$, every limit point of the sequence $\left\{\bar{\mathbf{V}}_s\right\}_{s\in\mathbb{N}}$ is an optimal solution of \eqref{pv1}.
\end{prop}
\begin{IEEEproof}
	Please refer to \cite[Appendix C]{9133130}.
\end{IEEEproof}
Now, the objective function of problem \eqref{pv2} is in a d.c. form, and thus, a stationary point of problem \eqref{pv2} can be obtained by applying SCA. In particular, in each iteration of the SCA algorithm, we construct a global  underestimator of the objective function by leveraging the following first-order Taylor approximation: 
\begin{equation}\label{eq15}
\left\Vert\bar{\mathbf{V}}\right\Vert_2\ge
\left\Vert\bar{\mathbf{V}}^{(t)}\right\Vert_2+\Tr\left[\boldsymbol{\lambda}_{\max}\left(\bar{\mathbf{V}}^{(t)}\right)\boldsymbol{\lambda}_{\max}^H\left(\bar{\mathbf{V}}^{(t)}\right)\left(\bar{\mathbf{V}}-\bar{\mathbf{V}}^{(t)}\right)\right],
\end{equation}
where  $\bar{\mathbf{V}}^{(t)}$ is the solution obtained in the $t$-th iteration.
Therefore, the optimization problem that needs to be solved in the $(t+1)$-th iteration is given by
\begin{equation}\label{pv3}
\begin{aligned}
&\underset{P>0,\bar{\mathbf{V}}\in\mathbb{H}^{M+1}\succeq\mathbf{0}}{\mathrm{minimize}} && P+\frac{1}{\mu}\left(\left\Vert\bar{\mathbf{V}}\right\Vert_*-\left\Vert\bar{\mathbf{V}}^{(t)}\right\Vert_2-\Tr\left[\boldsymbol{\lambda}_{\max}\left(\bar{\mathbf{V}}^{(t)}\right)\boldsymbol{\lambda}_{\max}^H\left(\bar{\mathbf{V}}^{(t)}\right)\left(\bar{\mathbf{V}}-\bar{\mathbf{V}}^{(t)}\right)\right]\right)\\
&\,\,\,\,\,\mathrm{subject\thinspace to}
&&\overline{\widehat{\mbox{C2}}}\mbox{:}\,\mathrm{Diag}\left(\bar{\mathbf{V}}\right)=P\mathbf{1}_{M+1},\quad\gamma_k\sigma_k^2+\Tr\left(\bar{\mathbf{V}}\mathbf{H}_k^H\tilde{\bar{\mathbf{W}}}_k\mathbf{H}_k\right)\le0,\quad\forall k,
\end{aligned}
\end{equation}
which is jointly convex with respect to $P$ and $\bar{\mathbf{V}}$ and therefore can be solved optimally via standard convex program solvers such as CVX \cite{grant2008cvx}.
\begin{algorithm}[t]
	\caption{SCA Algorithm}
	\begin{algorithmic}[1]
		\STATE Initialize $\bar{\mathbf{V}}^{(0)}$ with $M+1$ random phases. Set the convergence tolerance $0<\varepsilon\ll1$, penalty factor $0<\mu\ll1$, and iteration index $t=0$;
		\REPEAT 
		\STATE For a given $\bar{\mathbf{V}}^{(t)}$, update $\bar{\mathbf{V}}^{(t+1)}$ as the optimal solution of problem \eqref{pv3};
		\STATE $t\leftarrow t+1$;
		\UNTIL $\frac{M+1-\lambda_{\max}\left(\bar{\mathbf{V}}^{(t)}\right)}{M+1}\le\varepsilon$
	\end{algorithmic}
\end{algorithm}
\begin{algorithm}[t]
	\caption{Penalty-Based AltMin Algorithm}
	\begin{algorithmic}[1]
		\STATE Initialize $\mathbf{V}^{(0)}$ with $M+1$ random phases. Set the convergence tolerance  $0<\varepsilon\ll1$  and iteration index $t=0$;
		\REPEAT
		\STATE For a given  $\mathbf{V}^{(t)}$, update  ${\mathbf{W}}^{(t+1)}$ as the optimal solution of problem \eqref{p33};
		\STATE Set $P^{(t)}=\sum_{k\in\mathcal{K}}\Tr\left(\mathbf{W}_k^{(t+1)}\right)$ and $\bar{\mathbf{W}}_k^{(t+1)}={\mathbf{W}}_k^{(t+1)}/P^{(t)}$;
		\STATE For a given $P^{(t)}$ and $\bar{\mathbf{W}}_k^{(t+1)}$, update $P^{(t+1)}$ and $\bar{\mathbf{V}}^{(t+1)}$ by applying \textbf{Algorithm 1};
		\STATE Set $\mathbf{V}^{(t+1)}=\bar{\mathbf{V}}^{(t+1)}/P^{(t+1)}$;
		\STATE $t\leftarrow t+1$;
		\UNTIL $\frac{P^{(t-1)}-P^{(t)}}{P^{(t)}}\le\varepsilon$
	\end{algorithmic}
\end{algorithm}
The resulting SCA algorithm is  summarized in \textbf{Algorithm 1}.
By applying SCA, the minimum objective value of problem \eqref{pv3} serves as an upper bound for the optimal value of problem \eqref{pv2}.
By iteratively solving problem \eqref{pv3} optimally, we can monotonically tighten this upper bound. In this way, the objective values achieved by the sequence $\left\{P^{(t)},\bar{\mathbf{V}}^{(t)}\right\}_{t\in\mathbb{N}}$ form a non-increasing sequence that converges to a stationary point of problem \eqref{pv2} in polynomial time. 

The overall penalty-based AltMin algorithm is summarized in \textbf{Algorithm 2}. Basically, it has the structure of a two-block Gauss-Seidel  algorithm, where stationary points are obtained for both blocks defined by problems \eqref{pwk} and \eqref{pv1}, respectively. Therefore, the algorithm is guaranteed to converge to a stationary point of problem \eqref{refor} in polynomial time \cite{razaviyayn2013unified}.
Moreover, the computational complexity of each iteration of the proposed penalty-based AltMin algorithm is given by $\mathcal{O}\left(\ln\frac{1}{\rho}\left(K^2\left(\Nt M\right)^\frac{7}{2}+I\left(KM^\frac{7}{2}+K^2M^\frac{5}{2}\right)\right)\right)$, where  $\mathcal{O}\left(\cdot\right)$ is the big-O notation, $\rho$ is the convergence tolerance when applying the interior point method to solve an SDP problem, and  $I$ is the required number of iterations of Algorithm 1
\cite[Th. 3.12]{polik2010interior}.

\subsection{Proposed Inner Approximation (IA) Algorithm}

In addition to the penalty-based AltMin algorithm, we propose another novel algorithm to tackle \eqref{refor} by leveraging the IA technique. Instead of alternately optimizing the beamforming matrices $\mathbf{W}_k$ and the IRS reflecting elements in $\mathbf{V}$, we optimize these two  variables concurrently in each iteration of the IA algorithm, which avoids the feasibility problem \eqref{eq4} during AltMin.
Note that the non-convexity in problem \eqref{refor} resides in  constraints $\widehat{\mbox{C1}}$, \mbox{C3}, and \mbox{C4}. The core concept of IA is to  approximate in each iteration the non-convex feasible set by a convex set, which is more tractable for optimization. 
Next, we leverage the IA method to tackle the non-convex constraints $\widehat{\mbox{C1}}$ and \mbox{C3} in problem \eqref{refor}. 

The second term in the non-convex QoS constraint $\widehat{\mbox{C1}}$ can be rewritten as
\begin{equation}\label{eq10}
\Tr\left(\tilde{\mathbf{W}}_k\mathbf{H}_k\mathbf{V}\mathbf{H}_k^H\right)=\frac{1}{2}\left\Vert\tilde{\mathbf{W}}_k+\mathbf{H}_k\mathbf{V}\mathbf{H}_k^H\right\Vert_F^2-\frac{1}{2}\left\Vert\tilde{\mathbf{W}}_k\right\Vert_F^2
-\frac{1}{2}\left\Vert\mathbf{H}_k\mathbf{V}\mathbf{H}_k^H\right\Vert_F^2.
\end{equation}
Now, constraint $\widehat{\mbox{C1}}$ can be rewritten in form of a d.c. function, where the last two terms in \eqref{eq10} are convex with respect to $\mathbf{W}_k$ and $\mathbf{V}$, respectively.
To facilitate the application of  IA, we construct a global underestimator for the non-convex terms based on their first-order Taylor approximation. Specifically, we have 
\begin{equation}\label{eq12}
\begin{split}
\left\Vert\tilde{\mathbf{W}}_k\right\Vert_F^2&\ge-\left\Vert\tilde{\mathbf{W}}_k^{(t)}\right\Vert_F^2+2\Tr\left(\tilde{\mathbf{W}}_k^{(t)}\tilde{\mathbf{W}}_k\right),\\
\left\Vert\mathbf{H}_k\mathbf{V}\mathbf{H}_k^H\right\Vert_F^2&\ge-\left\Vert\mathbf{H}_k\mathbf{V}^{(t)}\mathbf{H}_k^H\right\Vert_F^2+2\Tr\left(\mathbf{H}_k^H\mathbf{H}_k\mathbf{V}^{(t)}\mathbf{H}_k^H\mathbf{H}_k\mathbf{V}\right),
\end{split}
\end{equation}
where $\tilde{\mathbf{W}}_k^{(t)}$ and $\mathbf{V}^{(t)}$ are the solutions obtained in the $t$-th iteration of IA.
Therefore, in the $(t+1)$-th iteration of the IA algorithm, the non-convex feasible set defined by $\widehat{\mbox{C1}}$ is approximated by
\begin{equation}
\begin{split}
\widetilde{\mbox{C1}}\mbox{:}\,&\frac{1}{2}\left\Vert\tilde{\mathbf{W}}_k+\mathbf{H}_k\mathbf{V}\mathbf{H}_k^H\right\Vert_F^2-\Tr\left(\mathbf{H}_k^H\mathbf{H}_k\mathbf{V}^{(t)}\mathbf{H}_k^H\mathbf{H}_k\mathbf{V}\right)-\Tr\left(\tilde{\mathbf{W}}_k^{(t)}\tilde{\mathbf{W}}_k\right)\\
&+\gamma_k\sigma_k^2+\frac{1}{2}\left\Vert\mathbf{H}_k\mathbf{V}^{(t)}\mathbf{H}_k^H\right\Vert_F^2+\frac{1}{2}\left\Vert\tilde{\mathbf{W}}_k^{(t)}\right\Vert_F^2
\le0,\quad\forall k,
\end{split}
\end{equation}
such that $\widetilde{\mbox{C1}}\Rightarrow\widehat{\mbox{C1}}$. In fact,
constraint $\widetilde{\mbox{C1}}$ is more tractable than $\widehat{\mbox{C1}}$, which facilitates the subsequent optimization and serves as the key step in deriving the proposed algorithm.

By employing the results in Lemma 1 and \eqref{eq15}, 
we obtain a convex approximation of  constraint \mbox{C3}, which is given by  
\begin{equation}
\widetilde{\mbox{C3}}\mbox{:}\,\left\Vert\mathbf{V}\right\Vert_*-\Tr\Big[\boldsymbol{\lambda}_{\max}\left(\mathbf{V}^{(t)}\right)\boldsymbol{\lambda}_{\max}^H\left(\mathbf{V}^{(t)}\right)\left(\mathbf{V}-\mathbf{V}^{(t)}\right)\Big]-\left\Vert\mathbf{V}^{(t)}\right\Vert_2\le0.
\end{equation}
In particular, we replace \mbox{C3} with $\widetilde{\mbox{C3}}$, which ensures that the rank-one constraint \mbox{C3} is satisfied after the convergence of the IA algorithm \cite{marks1978general}, which means that the diagonal elements of $\mathbf{\Phi}$, i.e., the first $M$ elements of $\mathbf{v}$, can always be recovered from the Cholesky decomposition of $\mathbf{V}=\mathbf{vv}^H$.

With the approximated convex constraints $\widetilde{\mbox{C1}}$ and $\widetilde{\mbox{C3}}$ at hand, the optimization problem  to be solved in the $(t+1)$-th iteration of the  IA algorithm is given by
\begin{equation}\label{overall}
\begin{aligned}
&\underset{\mathbf{W}_k\in\mathbb{H}^{\Nt},\mathbf{V}\in\mathbb{H}^{M+1}\succeq\mathbf{0}}{\mathrm{minimize}} &&f(\mathbf{W}_k)= \sum_{k\in\mathcal{K}}\Tr\left(\mathbf{W}_k\right)\\
&\mathrm{\quad\,\,\,\, subject\thinspace to}&&\widetilde{\mbox{C1}},\widehat{\mbox{C2}},\widetilde{\mbox{C3}},\mbox{C4}.
\end{aligned}
\end{equation}
We note that the remaining non-convexity of problem \eqref{overall} stems from the $K$ rank-one constraints in \mbox{C4}. To tackle this issue, we remove constraint \mbox{C4} by applying SDR. The resulting relaxed version of problem \eqref{overall} can be efficiently solved via standard convex program solvers such as CVX \cite{grant2008cvx}. Remarkably, unlike the SDR adopted for solving problem \eqref{eq5}, the tightness of this SDR can be guaranteed and is revealed in the following theorem.
\begin{thm}
	An optimal beamforming matrix $\mathbf{W}_k^\mathrm{opt}$ satisfying $\Rank\left(\mathbf{W}_k^\mathrm{opt}\right)\le1$ can always be obtained when solving problem \eqref{overall} without constraint \mbox{C4}.
\end{thm}
\begin{IEEEproof}
	Please refer to Appendix \ref{appA} for the proof.
\end{IEEEproof}

This theorem guarantees that the optimal beamforming vectors $\mathbf{w}_k^\mathrm{opt}$ in \eqref{overall} can always be recovered by performing Cholesky decomposition of the rank-one matrices $\mathbf{W}_k^\mathrm{opt}=\mathbf{w}_k^\mathrm{opt}\left(\mathbf{w}_k^\mathrm{opt}\right)^H$.
By employing the IA in \eqref{eq10} and \eqref{eq12}, the optimization variables in constraint \mbox{C1} are decoupled and therefore they can be optimized concurrently in each iteration of the proposed IA algorithm. This operation avoids the feasibility check problem \eqref{eq4}, which is the main advantage of our proposal compared to the existing approach\footnote{An intriguing observation in the practical implementation of the proposed IA algorithm is that the algorithm always generates a sequence of $\left\{\mathbf{V}^{(t)}\right\}_{t\in\mathbb{N}}$ that converges to a rank-one solution even if  constraint $\widetilde{\mbox{C3}}$ is dropped when solving problem \eqref{overall}. Nonetheless, the proof of this tightness is rather intricate and  beyond the scope of this paper.}.
The overall IA algorithm is summarized in \textbf{Algorithm 3}. According to \cite[Th. 1]{marks1978general}, the objective function $f$ in \eqref{refor} is non-increasing in each iteration and the proposed algorithm is guaranteed to converge to a KKT solution of problem \eqref{problem}.
The computational complexity of each iteration of the proposed IA algorithm is given by $\mathcal{O}\left(\log\frac{1}{\rho}\left(K\Nt^{\frac{7}{2}}+M^{\frac{7}{2}}\right)\right)$ \cite[Th. 3.12]{polik2010interior}. Note that while the KKT solution obtained by the IA algorithm is better in  quality than the stationary point obtained by the penalty-based AltMin algorithm, a trade-off between convergence rate and performance in terms of total transmit power minimization shall be revealed via simulations in Section \ref{VC}.
\begin{algorithm}[t]
	\caption{Inner Approximation (IA) Algorithm}
	\begin{algorithmic}[1]
		\STATE Initialize $\mathbf{V}^{(0)}$ with $M+1$ random phases and obtain $\mathbf{W}_k^{(0)}$ by solving problem \eqref{p33}. Set $0<\varepsilon\ll1$ and iteration index $t=0$;
		\REPEAT 
		\STATE For a given $\mathbf{W}_k^{(t)}$ and $\mathbf{V}^{(t)}$, update $\mathbf{W}_k^{(t+1)}$ and $\mathbf{V}^{(t+1)}$ as the optimal solution of problem \eqref{overall} without constraint \mbox{C4};
		\STATE $t\leftarrow t+1$;
		\UNTIL $\frac{f\left(\mathbf{W}_k^{(t-1)}\right)-f\left(\mathbf{W}_k^{(t)}\right)}{f\left(\mathbf{W}_k^{(t)}\right)}\le\varepsilon$
	\end{algorithmic}
\end{algorithm}

\section{Algorithm Design for IRS-Assisted  MISO Systems With  Imperfect CSI}
In this section, we first introduce the CSI uncertainty model and formulate a robust resource allocation problem for IRS-assisted multiuser MISO systems. Then, we develop an effective algorithm for jointly optimizing the beamformers at the AP and IRS reflecting elements.

\subsection{Channel State Information (CSI)}
While the direct links from the AP to user $k$, i.e., $\mathbf{d}_k$, can be estimated via conventional techniques, estimating  the reflected channels, namely $\mathbf{h}_k$ and $\mathbf{F}$, is more challenging.
In particular, IRSs are typically implemented by passive devices. Therefore, without the help of active RF chains, they are not able to receive and transmit signals to facilitate the channel estimation of the  reflected channels.
Instead, the pilot signals transmitted from the AP are reflected by the IRSs passively and are then received by the users. Therefore, the estimated CSI is the cascaded reflected channel, i.e., $\mathbf{E}_k=\mathrm{diag}\left(\mathbf{h}^H_k\right)\mathbf{F}$, rather than the two individual reflected channels themselves \cite{hu2019two,zhou2020framework,lin2020channel,9130088}.
In this paper, to account for the estimation error of the acquired CSI, we adopt a norm-bounded CSI error model \cite{6781609,zheng2008robust,4838902}.
Specifically, the CSI of both the cascaded reflected channels and the direct links is modeled as follows
\begin{equation}\label{uncertainty}
\begin{split}
\mathbf{E}_k&=\bar{\mathbf{E}}_k+\boldsymbol{\Delta}\mathbf{E}_k,\quad \Omega_{\mathbf{E}_k}\triangleq\left\{\boldsymbol{\Delta}\mathbf{E}_k\in\mathbb{C}^{M\times\Nt}:\left\Vert\boldsymbol{\Delta}\mathbf{E}_k\right\Vert_F\le\epsilon_{\mathbf{E}_k}\right\},\quad k\in\mathcal{K},\\
\mathbf{d}_k&=\bar{\mathbf{d}}_k+\boldsymbol{\Delta}\mathbf{d}_k,\quad \Omega_{\mathbf{d}_k}\triangleq\left\{\boldsymbol{\Delta}\mathbf{d}_k\in\mathbb{C}^{\Nt\times1}\left\Vert\boldsymbol{\Delta}\mathbf{d}_k\right\Vert_2\le\epsilon_{\mathbf{d}_k}\right\},\quad k\in\mathcal{K},
\end{split}
\end{equation}
where $\bar{\mathbf{E}}_k$ and $\bar{\mathbf{d}}_k$ are the estimates of the cascaded reflected channel and the direct channel between the AP and user $k$, respectively.
The corresponding CSI estimation errors are denoted by $\boldsymbol{\Delta}\mathbf{E}_k$ and $\boldsymbol{\Delta}\mathbf{d}_k$, whose norms are bounded by $\epsilon_{\mathbf{E}_k}$ and $\epsilon_{\mathbf{d}_k}$, respectively, where the sets $\Omega_{\mathbf{E}_k}$ and $\Omega_{\mathbf{d}_k}$ contain all possible  CSI estimation errors.
Note that the parameters $\epsilon_{\mathbf{E}_k}$ and $\epsilon_{\mathbf{d}_k}$ represent the level of CSI uncertainty, and are generally smaller for more accurate channel estimation and quantization algorithms.
The adopted model in \eqref{uncertainty} is flexible and general. In fact, it is able to capture various imperfections in the channel estimation process in IRS-assisted systems, e.g., noisy channel estimation \cite{9130088}, quantization errors of the phase shifts at the IRSs \cite{9133142}, and limited feedback \cite{8937491}. 

\subsection{Problem Formulation}
Next, we formulate a worst-case robust transmit power minimization problem for  norm-bounded CSI uncertainties. Referring to problem \eqref{refor} and \eqref{uncertainty}, the joint robust design of the beamformers at the AP and the phase shifts at the IRSs is formulated as follows
\begin{equation}\label{robustformulation}
\begin{aligned}
&\underset{\mathbf{W}_k\in\mathbb{H}^{\Nt},\mathbf{V}\in\mathbb{H}^{M+1}\succeq\mathbf{0}}{\mathrm{minimize}} && \sum_{k\in\mathcal{K}}\Tr\left(\mathbf{W}_k\right)\\
&\quad\,\,\,\,\mathrm{subject\thinspace to}
&&\widehat{\mbox{C2}}, \mbox{C3}, \mbox{C4},\\
&&&\mbox{C5}\mbox{:}\,\underset{\boldsymbol{\Delta}\mathbf{E}_k\in\Omega_{\mathbf{E}_k},\boldsymbol{\Delta}\mathbf{d}_k\in\Omega_{\mathbf{d}_k}}{\max}\gamma_k\sigma_k^2+\Tr\left(\mathbf{V}\mathbf{H}_k^H\tilde{\mathbf{W}}_k\mathbf{H}_k\right)\le0,\quad\forall k.
\end{aligned}
\end{equation}

The main difficulty in solving problem \eqref{robustformulation} lies in the rank-one constraint \mbox{C3} and constraint \mbox{C5}. In particular, constraint \mbox{C5} includes infinitely many  non-convex inequality constraints due to the continuity of the CSI uncertainty sets $\Omega_{\mathbf{E}_k}$ and $\Omega_{\mathbf{d}_k}$.
To address this difficulty, we convert constraint \mbox{C5} to  a finite number of  linear matrix inequality (LMI) constraints by leveraging the following lemma.
\begin{lem}\label{lem2}
	\emph{(S-Procedure \cite{boyd2004convex})} Let $f_i$, $i\in\{1,2\}$, be a real-valued function of vector $\mathbf{x}\in\mathbb{C}^{N\times1}$ and be defined as
	\begin{equation}
	f_i\left(\mathbf{x}\right)=\mathbf{x}^H\mathbf{A}_i\mathbf{x}+2\Re\left(\mathbf{a}^H_i\mathbf{x}\right)+a_i,
	\end{equation}
	where $\mathbf{A}_i\in\mathbb{H}^N$, $\mathbf{a}_i\in\mathbb{C}^{N\times1}$, and $a_i\in\mathbb{R}$. Then, the implication $f_1(\mathbf{x})\le0\Rightarrow f_2(\mathbf{x})\le0$ holds if and only if there exists a variable $q\ge0$ such that 
	\begin{equation}
	q\begin{bmatrix}
	\mathbf{A}_1&\mathbf{a}_1\\
	\mathbf{a}_1^H&a_1
	\end{bmatrix}-	\begin{bmatrix}
	\mathbf{A}_2&\mathbf{a}_2\\
	\mathbf{a}_2^H&a_2
	\end{bmatrix}\succeq\mathbf{0},
	\end{equation}
	provided that there exists a point $\bar{\mathbf{x}}$ that satisfies $f_i\left(\bar{\mathbf{x}}\right)<0$.
\end{lem}

Recall that we have defined the effective channel between the AP and user $k$ as
\begin{equation}
\mathbf{H}_k=\begin{bmatrix}
\mathbf{E}_k^H&\mathbf{d}_k
\end{bmatrix}=\begin{bmatrix}
\bar{\mathbf{E}}_k+\boldsymbol{\Delta}\mathbf{E}_k&\bar{\mathbf{d}}_k+\boldsymbol{\Delta}\mathbf{d}_k
\end{bmatrix}\triangleq\bar{\mathbf{H}}_k+\boldsymbol{\Delta}\mathbf{H}_k,
\end{equation}
where $\bar{\mathbf{H}}_k=\begin{bmatrix}
\mathbf{E}_k^H&\mathbf{d}_k
\end{bmatrix}$ and $\boldsymbol{\Delta}\mathbf{H}_k=\begin{bmatrix}
\boldsymbol{\Delta}\mathbf{E}_k^H&\boldsymbol{\Delta}\mathbf{d}_k
\end{bmatrix}$. According to \eqref{uncertainty}, we have
\begin{equation}
\left\Vert \boldsymbol{\Delta}\mathbf{H}_k\right\Vert_F=\sqrt{\left\Vert \boldsymbol{\Delta}\mathbf{E}_k\right\Vert_F^2+\left\Vert \boldsymbol{\Delta}\mathbf{d}_k\right\Vert_2^2}\le\sqrt{\epsilon_{\mathbf{E}_k}^2+\epsilon_{\mathbf{d}_k}^2}\triangleq\epsilon_k.
\end{equation} 
As $\Tr\left(\mathbf{A}^H\mathbf{BCD}\right)=\mathrm{vec}^H\left(\mathbf{A}\right)\left(\mathbf{D}^T\otimes\mathbf{B}\right)\mathrm{vec}\left(\mathbf{C}\right)$, we can rewrite constraint \mbox{C5} as follows
\begin{equation}\label{eq24}
\gamma_k\sigma_k^2+\Tr\left(\mathbf{V}\mathbf{H}_k^H\tilde{\mathbf{W}}_k\mathbf{H}_k\right)=\gamma_k\sigma_k^2+\mathbf{g}^H_k\left(\mathbf{V}^T\otimes\tilde{\mathbf{W}}_k\right)\mathbf{g}_k\le0,\quad\forall \left\Vert \boldsymbol{\Delta}\mathbf{H}_k\right\Vert_F\le \epsilon_k,
\end{equation}
where $\mathbf{g}_k=\mathrm{vec}\left(\mathbf{H}_k\right)$. We further define  
\begin{equation}\label{eq25}
\mathbf{g}_k=\bar{\mathbf{g}}_k+\boldsymbol{\Delta}\mathbf{g}_k,
\end{equation} 
where $\bar{\mathbf{g}}_k=\mathrm{vec}\left(\bar{\mathbf{H}}_k\right)$, $\boldsymbol{\Delta}\mathbf{g}_k=\mathrm{vec}\left(\boldsymbol{\Delta}\mathbf{H}_k\right)$, and thus $\left\Vert \boldsymbol{\Delta}\mathbf{g}_k\right\Vert_2=\left\Vert \boldsymbol{\Delta}\mathbf{H}_k\right\Vert_F\le\epsilon_k$. By substituting \eqref{eq25} into \eqref{eq24}, constraint \mbox{C5} can be further recast as follows
\begin{equation}
\begin{split}
\mbox{C5}&\Leftrightarrow\boldsymbol{\Delta}\mathbf{g}_k^H\left(\mathbf{V}^T\otimes\tilde{\mathbf{W}}_k\right)\boldsymbol{\Delta}\mathbf{g}_k+2\Re\left(\bar{\mathbf{g}}_k^H\left(\mathbf{V}^T\otimes\tilde{\mathbf{W}}_k\right)\boldsymbol{\Delta}\mathbf{g}_k\right)\\
&\relphantom{\Rightarrow}+\bar{\mathbf{g}}_k^H\left(\mathbf{V}^T\otimes\tilde{\mathbf{W}}_k\right)\bar{\mathbf{g}}_k+\gamma_k\sigma_k^2\le0,\quad\forall\left\Vert \boldsymbol{\Delta}\mathbf{g}_k\right\Vert_2\le\epsilon_k.
\end{split}
\end{equation}
Then, by applying Lemma \ref{lem2}, constraint \mbox{C5} is rewritten as
\begin{equation}\label{eq27}
\begin{split}
\mbox{C5}&\Leftrightarrow
q_k\begin{bmatrix}
\mathbf{I}_{N_\mathrm{t}(M+1)}&\mathbf{0}\\
\mathbf{0}&-\epsilon_k^2
\end{bmatrix}-
\begin{bmatrix}
\mathbf{V}^T\otimes\tilde{\mathbf{W}}_k&\left(\mathbf{V}^T\otimes\tilde{\mathbf{W}}_k\right)\bar{\mathbf{g}}_k\\
\bar{\mathbf{g}}_k^H\left(\mathbf{V}^T\otimes\tilde{\mathbf{W}}_k\right)&\bar{\mathbf{g}}_k^H\left(\mathbf{V}^T\otimes\tilde{\mathbf{W}}_k\right)\bar{\mathbf{g}}_k+\gamma_k\sigma_k^2
\end{bmatrix}\succeq\mathbf{0}\\
&\Leftrightarrow \overline{\mbox{C5}}\mbox{:}\,\mathbf{P}_k-\mathbf{G}_k^H\left(\mathbf{V}^T\otimes\tilde{\mathbf{W}}_k\right)\mathbf{G}_k\succeq\mathbf{0},\quad\forall k,
\end{split}
\end{equation}
where 
\begin{equation}\label{eq28}
\mathbf{P}_k=
\begin{bmatrix}
q_k\mathbf{I}_{N_\mathrm{t}(M+1)}&\mathbf{0}\\
\mathbf{0}&-q_k\epsilon_k^2-\gamma_k\sigma_k^2
\end{bmatrix},\quad\mathbf{G}_k=\begin{bmatrix}
\mathbf{I}_{N_\mathrm{t}(M+1)}&\bar{\mathbf{g}}_k
\end{bmatrix},
\end{equation}
 and $q_k\ge0$, $\forall k$. In \eqref{eq27},  the infinitely many inequalities in \mbox{C5} are transformed to $K$ LMIs in constraint $\overline{\mbox{C5}}$, which is preferable  for algorithm design. Hence, the original robust design problem \eqref{robustformulation} is rewritten as
 \begin{equation}\label{p29}
 \begin{aligned}
 &\underset{\mathbf{W}_k\in\mathbb{H}^{\Nt},\mathbf{V}\in\mathbb{H}^{M+1}\succeq\mathbf{0}}{\mathrm{minimize}} && \sum_{k\in\mathcal{K}}\Tr\left(\mathbf{W}_k\right)\\
 &\quad\,\,\,\,\mathrm{subject\thinspace to}
 &&\widehat{\mbox{C2}}, \mbox{C3}, \mbox{C4}, \overline{\mbox{C5}}.
 \end{aligned}
 \end{equation}
 
Unlike constraint $\widehat{\mbox{C1}}$ for the perfect CSI scenario, constraint $\overline{\mbox{C5}}$ is defined by a series of LMIs rather than  real-valued inequalities. Therefore, it is difficult to construct convex approximations of the feasible sets defined by $\overline{\mbox{C5}}$ and to apply the IA method to handle the robust design of IRS-assisted wireless systems. Hence, in this section, we extend the penalty-based AltMin algorithm in \textbf{Algorithm 1} to  problem \eqref{p29}.

\emph{Remark 3:} Unlike existing works \cite{9110587,9133130,hong2020robust,zhao2020outage} which assumed that only a part of the CSI is imperfect, typically only the CSI of the reflected channels,  we take into account the CSI uncertainties of all communication links in this paper.
While the imperfect CSI of the direct links was investigated  in \cite{zhang2020robust,xu2020resource,zhou2020framework}, in these works approximations or bounds  for constraint \mbox{C5} were used, which led to a safe approximation of the robust constraint. However, this may lead to a more restricted feasible set for optimization and may result in a higher total transmit power. To the best of the author's knowledge, the proposed equivalent transform of  constraint \mbox{C5} to $\overline{\mbox{C5}}$ is the first method in the literature that can accurately capture the impact of the imperfect CSI of all channels for robust optimization of IRS-assisted wireless systems.

\subsection{Proposed  Penalty-Based AltMin Algorithm}
Following a similar derivation procedure as from \eqref{p30} to \eqref{pv3},  two problems  need to be solved during the proposed AltMin optimization. 
Specifically, the optimization of $P$ and $\bar{\mathbf{W}}_k$ for a fixed $\mathbf{V}$ is given by
\begin{equation}\label{p332}
\underset{\mathbf{W}_k\in\mathbb{H}^{\Nt}\succeq\mathbf{0}}{\mathrm{minimize}} \,\,\,\, \sum_{k\in\mathcal{K}}\Tr\left(\mathbf{W}_k\right)\quad\quad\mathrm{subject\thinspace to}\,\,\,\,\mbox{C4}, \overline{\mbox{C5}},
\end{equation}
where $P=\sum_{k\in\mathcal{K}}\Tr\left(\mathbf{W}_k\right)$ and $\bar{\mathbf{W}}_k=\frac{1}{{P}}\mathbf{W}_k$. 
The tightness of the relaxation of the rank-one constraint \mbox{C4} is presented in the following theorem.
\begin{thm}\label{th2}
	An optimal beamforming matrix $\mathbf{W}_k^\mathrm{opt}$ satisfying $\Rank\left(\mathbf{W}_k^\mathrm{opt}\right)\le1$ can always be obtained for problem \eqref{p332} without constraint \mbox{C4}.
\end{thm}
\begin{IEEEproof}
	See Appendix \ref{appB}.
\end{IEEEproof}

On the other hand, when $\bar{\mathbf{W}}_k$ is given, the  problem that needs to be solved in the $(t+1)$-th iteration for optimizing $P$ and  $\bar{\mathbf{V}}$ is given by
\begin{equation}\label{pv32}
\begin{aligned}
&\underset{P>0,\bar{\mathbf{V}}\in\mathbb{H}^{M+1}\succeq\mathbf{0}}{\mathrm{minimize}} && P+\frac{1}{\mu}\left(\left\Vert\bar{\mathbf{V}}\right\Vert_*-\left\Vert\bar{\mathbf{V}}^{(t)}\right\Vert_2-\Tr\left[\boldsymbol{\lambda}_{\max}\left(\bar{\mathbf{V}}^{(t)}\right)\boldsymbol{\lambda}_{\max}^H\left(\bar{\mathbf{V}}^{(t)}\right)\left(\bar{\mathbf{V}}-\bar{\mathbf{V}}^{(t)}\right)\right]\right)\\
&\,\,\,\,\,\mathrm{subject\thinspace to}
&&\overline{\widehat{\mbox{C2}}},\quad\mathbf{P}_k-\mathbf{G}_k^H\left(\bar{\mathbf{V}}^T\otimes \tilde{\bar{\mathbf{W}}}_k\right)\mathbf{G}_k\succeq\mathbf{0},\quad\forall k.
\end{aligned}
\end{equation}
The resulting SCA and penalty-based AltMin algorithms can be summarized in a similar manner as \textbf{Algorithm 1} and \textbf{Algorithm 2}, respectively. In particular, problems \eqref{pv3} and \eqref{p33} are replaced  by problems  \eqref{pv32} and \eqref{p332}, respectively. The analysis of the convergence and computational complexity are identical as those  at the end of Section \ref{IIIB}.

\section{Simulation Results}
\begin{table}[t]
	\centering
	\begin{minipage}[t]{0.48\linewidth} 
		\centering\includegraphics[height=5cm]{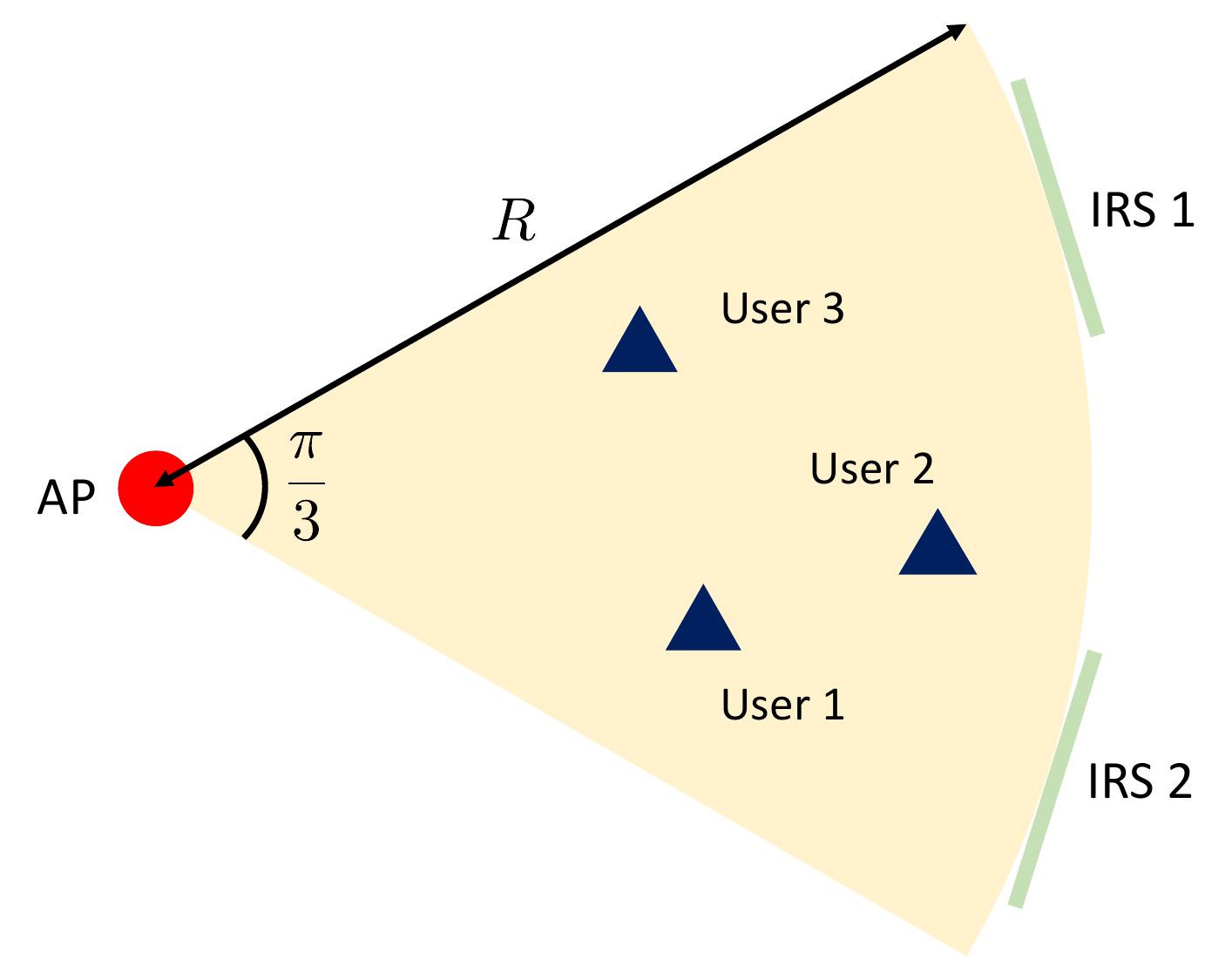}
		\captionof{figure}{Simulation setup for an IRS-assisted multiuser MISO  system  comprising $K = 3$ users and $L = 2$ IRSs.}
		\label{simulation}
	\end{minipage}\quad
	\begin{minipage}[t]{0.48\linewidth}\vspace{-15em} 
		\caption{Simulation parameters}
		\centering
		\begin{tabular}{|l||l|l|}
			\hline
			$f_\mathrm{c}$&Carrier center    frequency&$2.4$ GHz\\\hline
			$\sigma^2$&Noise power &$-90$ dBm\\\hline
			\multirow{2}{*}{$\alpha_\mathrm{L}$}&
			Path loss exponent for  &\multirow{2}{*}{$2.1$}\\
			&Ricean fading channels &\\\hline
			$\beta$&Ricean factor&$1$\\\hline
			\multirow{2}{*}{$\alpha_\mathrm{N}$}&
			Path loss exponent for  &\multirow{2}{*}{$4$}\\
			&Rayleigh fading channels &\\\hline
			$\varepsilon$&Convergence tolerance&$10^{-5}$\\\hline
		\end{tabular}
	\end{minipage}
\end{table}

\subsection{Simulation Setup}
The schematic system model adopted for the simulated IRS-assisted multiuser MISO system is shown in Fig. \ref{simulation}.
The AP serves one sector of a cell with radius $R$.  $K$ users are randomly and uniformly distributed in the sector and the IRSs are deployed at the edge of the cell.
The channel matrix $\mathbf{F}_l$ between the AP and IRS $l$ is modeled as
\begin{equation}
\mathbf{F}_l=\sqrt{L_0d_l^{-\alpha_\mathrm{L}}}\left(\sqrt{\frac{\beta}{1+\beta}}\mathbf{F}_l^\mathrm{L}+\sqrt{\frac{1}{1+\beta}}\mathbf{F}^\mathrm{N}_l\right),
\end{equation}
where $L_0=\left(\frac{\lambda_{c}}{4\pi}\right)^2$ is a constant with $\lambda_{c}$ being the wavelength of the carrier frequency. The distance between the AP and IRS $l$ is denoted by $d_l$ and $\alpha_\mathrm{L}$ is the path loss exponent. 
The small-scale fading is assumed to be Ricean fading with Ricean factor $\beta$. $\mathbf{F}_l^\mathrm{L}$ and $\mathbf{F}_l^\mathrm{N}$ are the line-of-sight (LoS) and non-LoS components, respectively. 
The LoS component is the product of the receive and transmit array response vectors while the non-LoS component is modeled by Rayleigh fading. The channel vectors $\mathbf{h}_{kl}$ between IRS $l$ and user $k$ are generated in a similar manner as $\mathbf{F}_l$. 
In addition, the direct links $\mathbf{d}_k$ between the AP and the users are modeled as pure non-LoS channels, i.e., Rayleigh fading,  since one of the motivations for deploying IRSs is that the direct links are shadowed by obstacles. 
The path loss exponents of these direct links are denoted by  $\alpha_\mathrm{N}$.
For the ease of presentation, we assume that the minimum required SINRs and noise powers at all users are identical, respectively, i.e., $\sigma^2_k=\sigma^2$,  $\gamma_k=\gamma$, $\forall k$, and define  maximum normalized estimation error of the channels as $\kappa_k=\epsilon_k/\left\Vert\bar{\mathbf{H}}_k\right\Vert_F=\kappa$, $\forall k$.
The important system parameters adopted in our simulations are listed in Table I.

\subsection{Baseline Schemes}
To show the effectiveness of the proposed algorithms in this paper, we adopt the state-of-the-art SDR-AltMin algorithm as one of the benchmarks. 
Since the SDR-based AltMin algorithm cannot guarantee convergence, for a fair comparison in Sections IV-D to IV-G, we set the maximum number of iterations equal to the number of iterations required by the proposed algorithms to converge.
In addition, if  QoS constraint $\widecheck{\mbox{C1}}$ is not satisfied by the normalized random vector $\mathbf{v}$, we add an additional step at the end of the SDR-AltMin algorithm, which solves problem \eqref{p1} with a fixed $\mathbf{v}$ to yield a feasible solution. 
Also, we consider two additional baseline schemes.
For baseline 1, we evaluate the performance of a conventional multiuser MISO system without the deployment of IRSs.
In particular, we optimize the beamformers for $\mathbf{\Phi}=\mathbf{0}$ and solve problems \eqref{p1} and \eqref{p33} for the case of perfect and imperfect CSI, respectively. 
For baseline 2, we adopt IRSs employing random phase shifts and optimize the
beamformers at the AP by solving problems \eqref{p1} and \eqref{p33}.
In the following, we investigate the impact of the different system parameters by focusing on the case of one IRS while multi-IRS systems are considered in Section \ref{VE}.

\subsection{Convergence of the Proposed Algorithms}\label{VC}
\begin{figure}[t]
	\centering
	\begin{minipage}[t]{0.485\linewidth} \hspace{-1em}
		\centering\includegraphics[height=6cm]{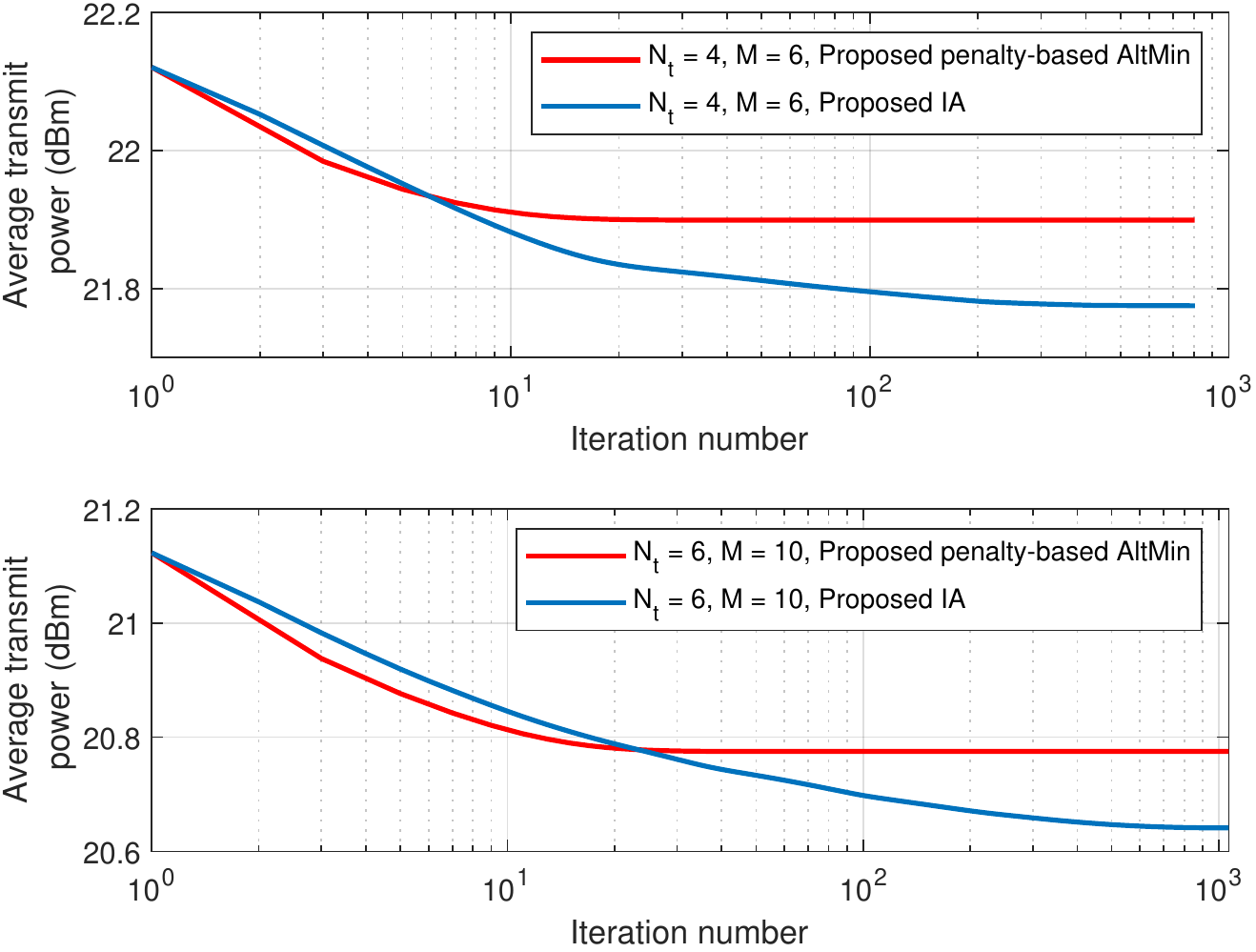}
		\caption{Convergence of the proposed penalty-based AltMin algorithm and IA algorithm for different values of $\Nt$ and $M$ with perfect CSI. The system parameters are set as $R=120$ m, $K=4$, and $\gamma=2$ dB.}\label{iteration1}
	\end{minipage}\quad
	\begin{minipage}[t]{0.485\linewidth} \hspace{-1em}
		\centering\includegraphics[height=6cm]{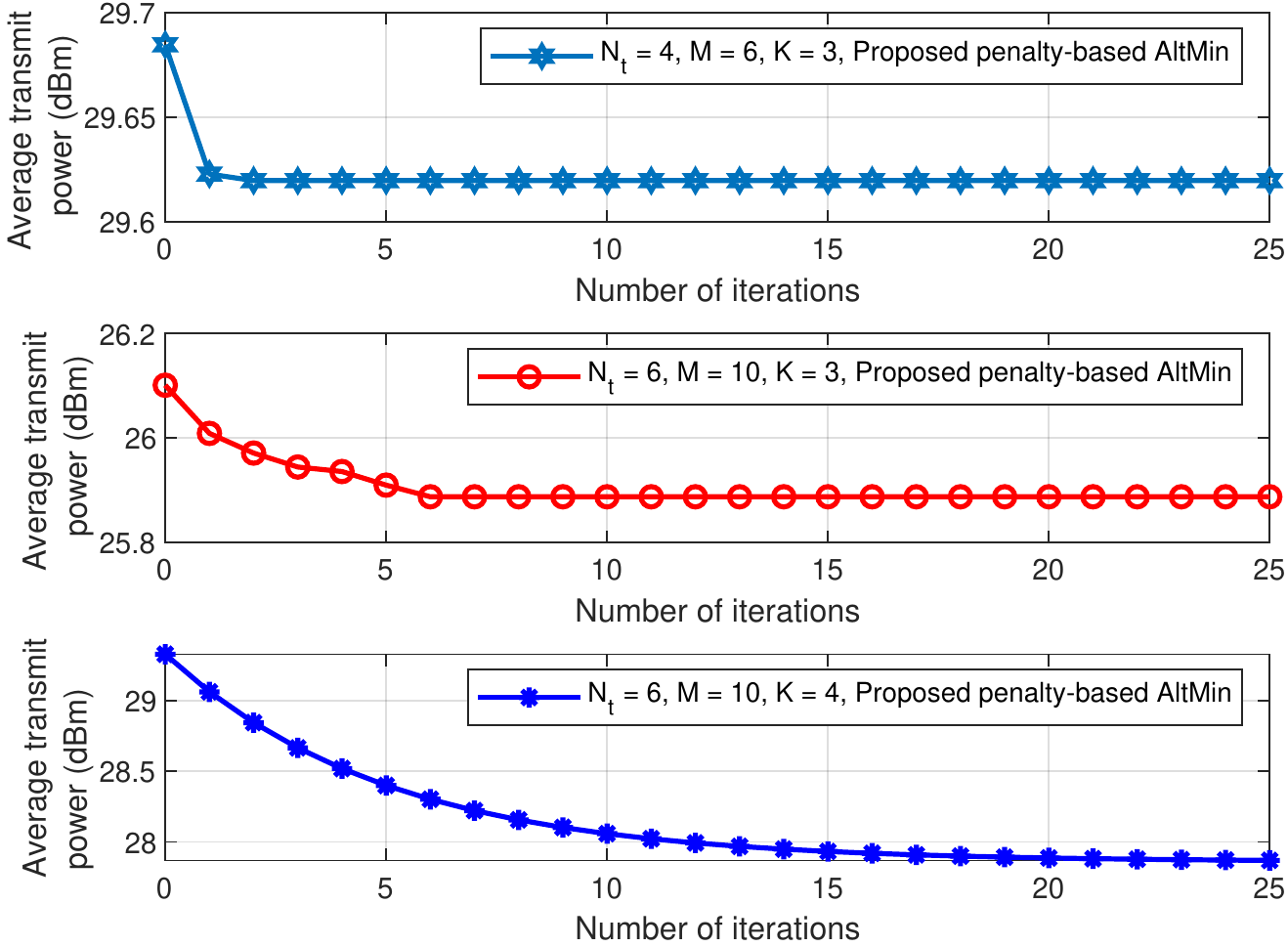}
		\caption{Convergence of the proposed penalty-based AltMin algorithm for different values of $\Nt$, $M$, and $K$ with imperfect CSI. The system parameters are set as $R=120$ m, $\gamma=2$ dB, and $\kappa=0.2$.} \label{iteration2}
	\end{minipage}
\end{figure}


We investigate the convergence of the proposed penalty-based AltMin  and IA algorithms  for different numbers of transmit antennas at the AP, $\Nt$, numbers of reflecting elements at IRS, $M$, and numbers of users, $K$. 
First, we compare the performance of the two proposed algorithms for the perfect CSI case in Fig. \ref{iteration1}. 
As can be observed, the penalty-based AltMin algorithm requires much fewer iterations to converge than the IA algorithm on average. In particular, when $\Nt=4$ and $M=6$, the proposed penalty-based AltMin  and IA algorithms converge within 15 and 300 iterations on average, respectively. Note that the QoS constraint $\widehat{\mbox{C1}}$ in problem \eqref{refor} is approximated in each iteration of the IA algorithm while it remains unchanged in the penalty-based AltMin algorithm. Therefore, the IA algorithm needs more iterations to converge to satisfy constraint $\widehat{\mbox{C1}}$.
In addition, for the case with more AP antenna elements and IRS reflecting elements, i.e., $\Nt=6$ and $M=10$, the two proposed algorithms converge within 30 and 800 iterations on average, respectively. This is because the dimensions of the solution space of problems \eqref{p33},  \eqref{pv3}, and \eqref{overall}  scale with $\Nt$ and $M$.
Furthermore, it is noted that the average transmit power achieved by the IA algorithm is lower than that of the penalty-based AltMin algorithm. This observation confirms that the IA algorithm converges to a KKT solution, which is better in quality than the stationary point obtained by the penalty-based AltMin algorithm. In summary, Fig. \ref{iteration1} clearly illustrates a trade-off between the convergence speed and performance in terms of total transmit power minimization.

Second, as expected, in Fig. \ref{iteration2}, the average transmit power of the proposed penalty-based AltMin algorithm monotonically decreases for all considered values of $\Nt$, $M$, and $K$ for imperfect CSI.
Similar to the observations in Fig. \ref{iteration1}, the proposed penalty-based AltMin algorithm on average requires more iterations to converge when the sizes of the antenna array and the IRS increase, as is evident from comparing the first two subfigures in Fig. \ref{iteration2}. Moreover,
for the case with more users, i.e., $\Nt=6$, $M=10$, and $K=4$, the penalty-based AltMin algorithm needs considerably more iterations (around 20 iterations) to converge since the solution space of problems \eqref{p332} and \eqref{pv32} increases with the number of users, $K$.

\subsection{Average Transmit Power Versus the Minimum Required SINR}\label{VE}

\begin{figure}[t]
	\centering
	\begin{minipage}[t]{0.485\linewidth} \hspace{-1em}
		\centering\includegraphics[height=6cm]{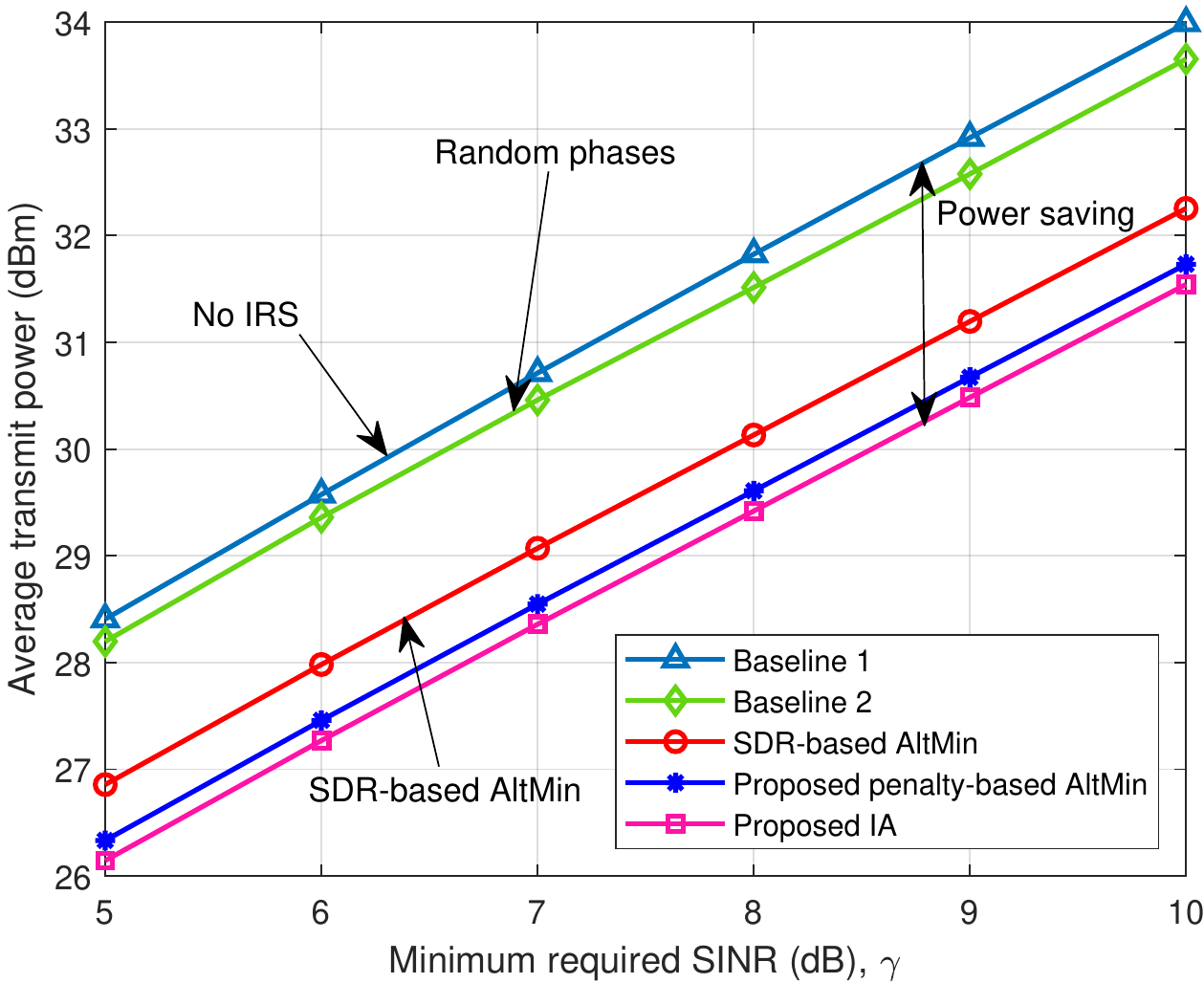}
		\caption{Average transmit power achieved by different algorithms when $R=100$ m, $\Nt=4$,  $M=30$, and $K=3$ with perfect CSI.} \label{fig2}
	\end{minipage}\quad
	\begin{minipage}[t]{0.485\linewidth} \hspace{-1em}
		\centering\includegraphics[height=6cm]{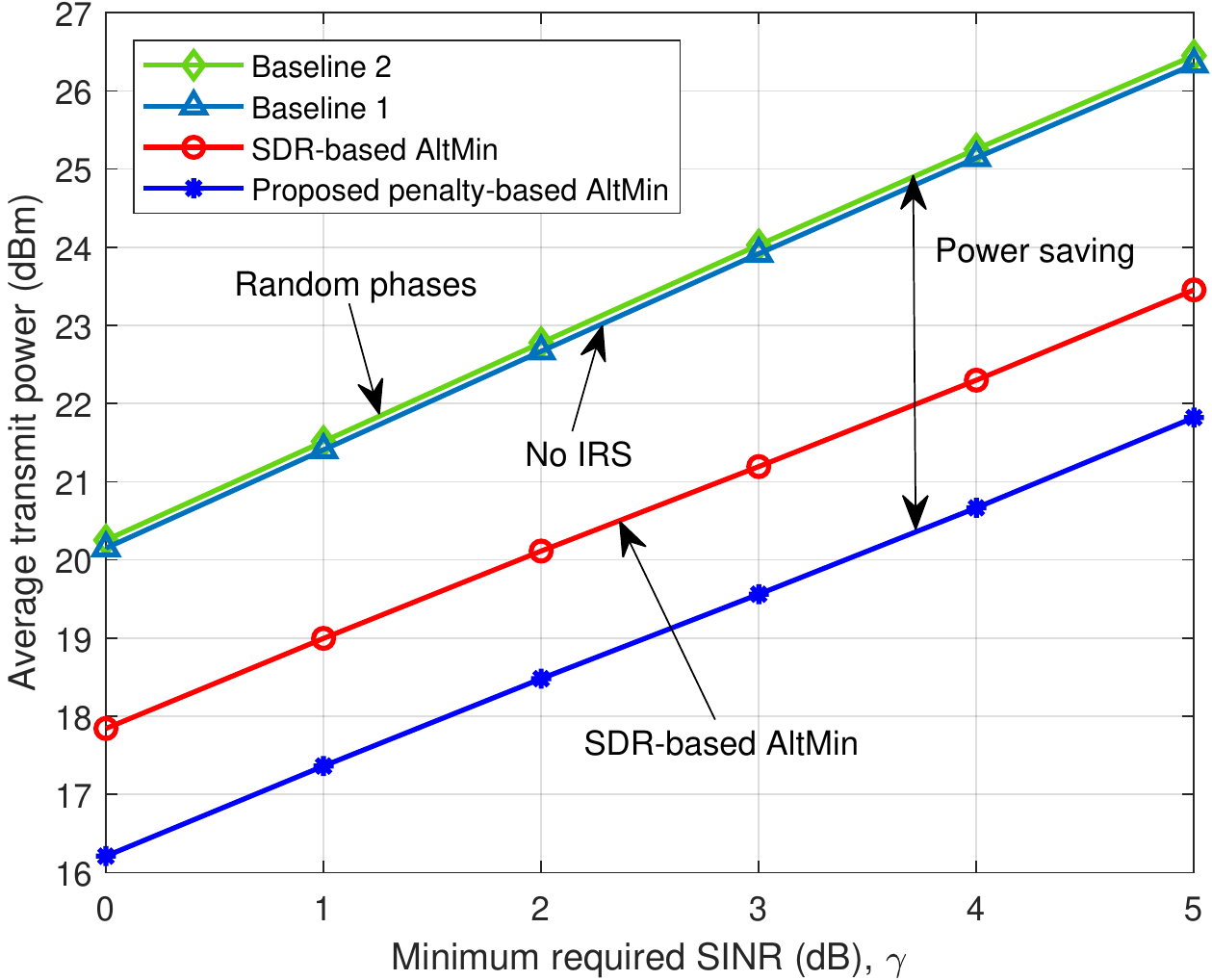}
		\caption{Average transmit power achieved by different algorithms when  $R=100$ m, $\Nt=4$,  $M_1=M_2=10$,  $K=3$, and $\kappa=0.05$.} \label{fig4}
	\end{minipage}
\end{figure}

In Fig. \ref{fig2}, the average transmit power of the AP is plotted for different minimum required SINR values, $\gamma$, when perfect CSI is available. 
As can be observed, the average transmit power  monotonically increases with the minimum required SINR, since more power is needed for satisfying more stringent QoS requirements.
We also note that deploying an IRS significantly reduces the required average transmit power in the considered multiuser MISO system. 
This shows the ability of IRSs to establish favorable channel conditions, which allow the system to guarantee the QoS of the users for lower transmit powers. Hence, deploying IRSs is a promising approach for enabling green wireless communication systems.
In addition, we observe that the two  proposed algorithms outperform the baseline scheme with random phase shifts, which reveals the necessity to carefully design the reflecting elements in IRS-assisted green wireless systems.
Finally, the two proposed  algorithms lead to a lower transmit power than the SDR-based AltMin algorithm, which cannot guarantee convergence.
These results clearly show the effectiveness of the two proposed  algorithms in jointly optimizing the beamformers at the AP and the IRS reflecting elements and guaranteeing  convergence.
On the other hand, we evaluate the average transmit power when two IRSs are deployed in the network, as shown in Fig. \ref{simulation}. In particular, we assume that  $M_1=M_2=10$ reflecting elements are available at each IRS.
Fig. \ref{fig4} shows the average transmit power versus the minimum required SINR, $\gamma$, when the CSI of all channels is not perfectly known at the AP. Similar to Fig. \ref{fig2}, the proposed penalty-based AltMin algorithm can reduce the total transmit power significantly compared to the three baseline schemes.
This again illustrates the benefits brought by deploying IRSs and the proposed penalty-based AltMin algorithm for robust and green wireless communication systems.
As can be observed, different from the perfect CSI scenario shown in Fig. \ref{fig2}, baseline scheme 1  achieves a lower transmit power than baseline scheme 2.
Indeed, when there are no IRSs deployed in the considered network (baseline 1), the robust resource allocation design has to combat only the CSI uncertainty of the direct links from the AP to the users. In contrast, in baseline 2, the imperfect CSI of both reflected channels also has to be taken into account for robust optimization. Therefore, more transmit power is required for robust resource allocation to combat additional channel uncertainties in baseline scheme 2.
This also indicates that when the CSI cannot be accurately acquired, both the reflecting elements at the IRSs and the beamformers at the AP have to be jointly optimized to achieve transmit power reduction.
Moreover,  the performance gap between the SDR-based AltMin algorithm and the proposed scheme is larger in Fig. \ref{fig4} than  in Fig. \ref{fig2}.
In particular,  the  imperfect CSI  reduces the size of the feasible solution set of the resource allocation problem. As such, the  solution obtained via Gaussian randomization falls in the feasible set with a much lower probability, which makes it more difficult to satisfy the robust QoS constraint \mbox{C5} via the SDR approach. Therefore, a higher transmit power is needed to compensate the loss in QoS while guaranteeing  robustness.

\subsection{CSI Uncertainty}\label{VD}
\begin{figure}[t]
	\centering
	\begin{minipage}[t]{0.485\linewidth} \hspace{-1em}
		\centering\includegraphics[height=6cm]{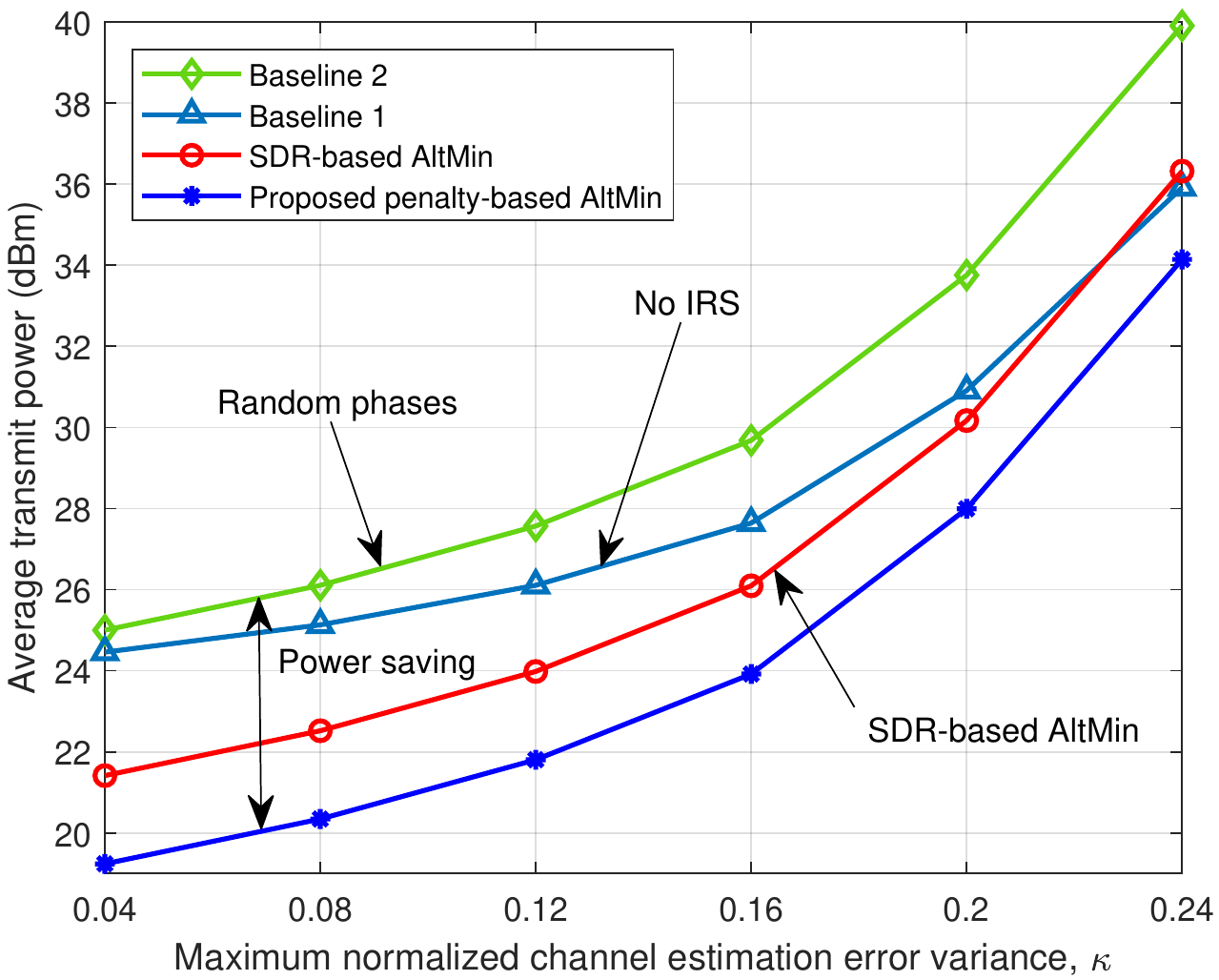}
		\caption{Average transmit power versus the maximum normalized channel estimation error variance when $R=80$ m, $\Nt=4$,  $M=10$,  $K=3$, and $\gamma=5$ dB.} \label{fig5}
	\end{minipage}\quad
	\begin{minipage}[t]{0.485\linewidth} \hspace{-1em}
		\centering\includegraphics[height=6cm]{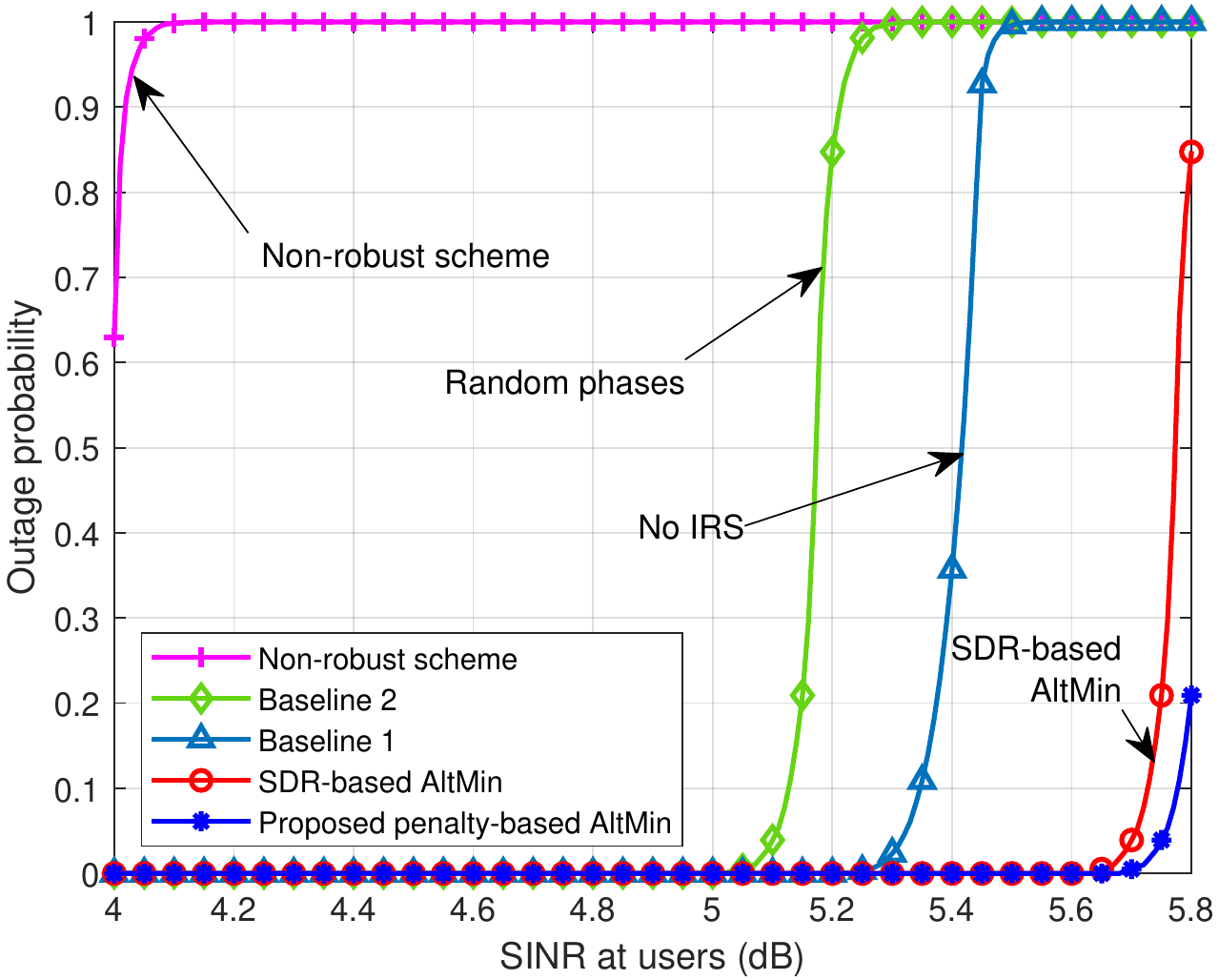}
		\caption{Outage probability of the users versus SINR value. The system parameters are set as $R=100$ m, $\Nt=M=10$,  $K=3$, $\gamma=5$ dB, and $\kappa=0.1$.} \label{fig6}
	\end{minipage}
\end{figure}
Fig. \ref{fig5} shows the average transmit power versus the maximum normalized channel estimation error variance, $\kappa$.
As can be observed, the average transmit power increases with the severity of the CSI degradation for both the proposed and the baseline schemes. In particular, the worse the quality of the estimated CSI is,  the more difficult it is for the AP to perform accurate beamforming to satisfy the QoS. Therefore, more transmit power is required to maintain the robustness of the system.
Note that the proposed penalty-based AltMin algorithm significantly outperforms the baseline schemes in terms of average transmit power, which illustrates that the proposed scheme is able to exploit the spatial degrees of freedom (DoFs) more effectively than existing approaches even in the presence of CSI uncertainty.
Furthermore, the performance gain achieved by the proposed scheme over baseline scheme 1 decreases with the maximum normalized channel estimation error variance, $\kappa$. In particular,  when no IRSs are deployed (baseline 1) the imperfect CSI only needs to be taken into account for the direct links while additional power needs to be transmitted to combat the CSI uncertainty for both the cascaded reflected channels and the direct links in the IRS-assisted system (baseline scheme 2, SDR-based AltMin, and  proposed schemes). This phenomenon is more prominent  when the estimation error variance is large. Therefore, as the CSI quality degrades, the average transmit power required by baseline scheme 1 grows slower than that of all other schemes designed for IRS-assisted systems.


In Fig. \ref{fig6}, the outage probability of the users versus the SINR threshold is plotted when $\kappa=0.1$.
The outage probability is defined as the probability that the received SINRs at the users are lower than a predefined  SINR threshold. 
In this subsection, we further investigate the performance of another scheme, namely a non-robust scheme, for comparison with the proposed robust design. In particular, for the non-robust scheme, the estimated channels $\bar{\mathbf{H}}_j$ are treated as  perfect CSI and the IA algorithm is applied to jointly optimize the beamformers and IRS reflecting elements.
First, because we set the minimum required SINR at the users to $\gamma=5$ dB, the outage probabilities of the proposed scheme and the three baseline schemes are zero when the  SINR threshold is no larger than $5$ dB. In contrast, the outage probability of the non-robust scheme at $5$ dB is one, which underscores the importance of the  proposed design approach.
More importantly, the outage probability of the proposed scheme is significantly lower than those of the three baselines even when the  SINR threshold is larger than  $5$ dB. In particular, for a  SINR threshold of $5.8$ dB, the outage probabilities of baseline schemes 1 and 2 are one while the outage probability of the SDR-based AltMin is $85$\%. In contrast, the proposed scheme can remarkably reduce the outage probability to  $20$\%.
This demonstrates that compared to the baseline schemes,  the proposed scheme is more capable of achieving higher user SINRs in the presence of CSI uncertainty. 

\subsection{  System Power Consumption Minimization}

\begin{figure}[t]
	\centering
	\begin{minipage}[t]{0.485\linewidth} \hspace{-1em}
		\centering\includegraphics[height=6cm]{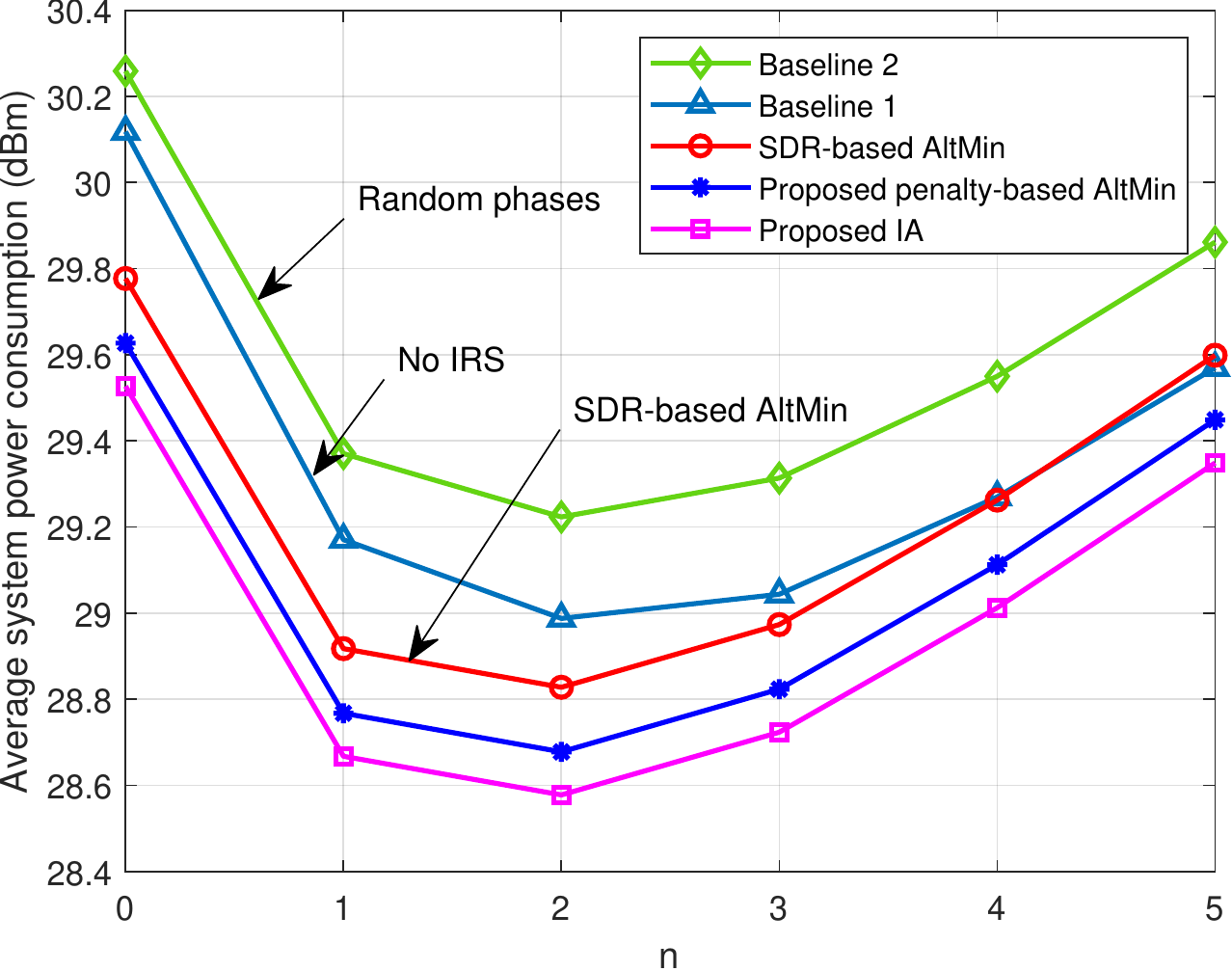}
		\caption{Average system power consumption achieved by different algorithms when $R=80$ m, $\Nt=K=4$,  $M=8$, and $\gamma=2$ dB with perfect CSI.} \label{fig8}
	\end{minipage}\quad
	\begin{minipage}[t]{0.485\linewidth} \hspace{-1em}
		\centering\includegraphics[height=6cm]{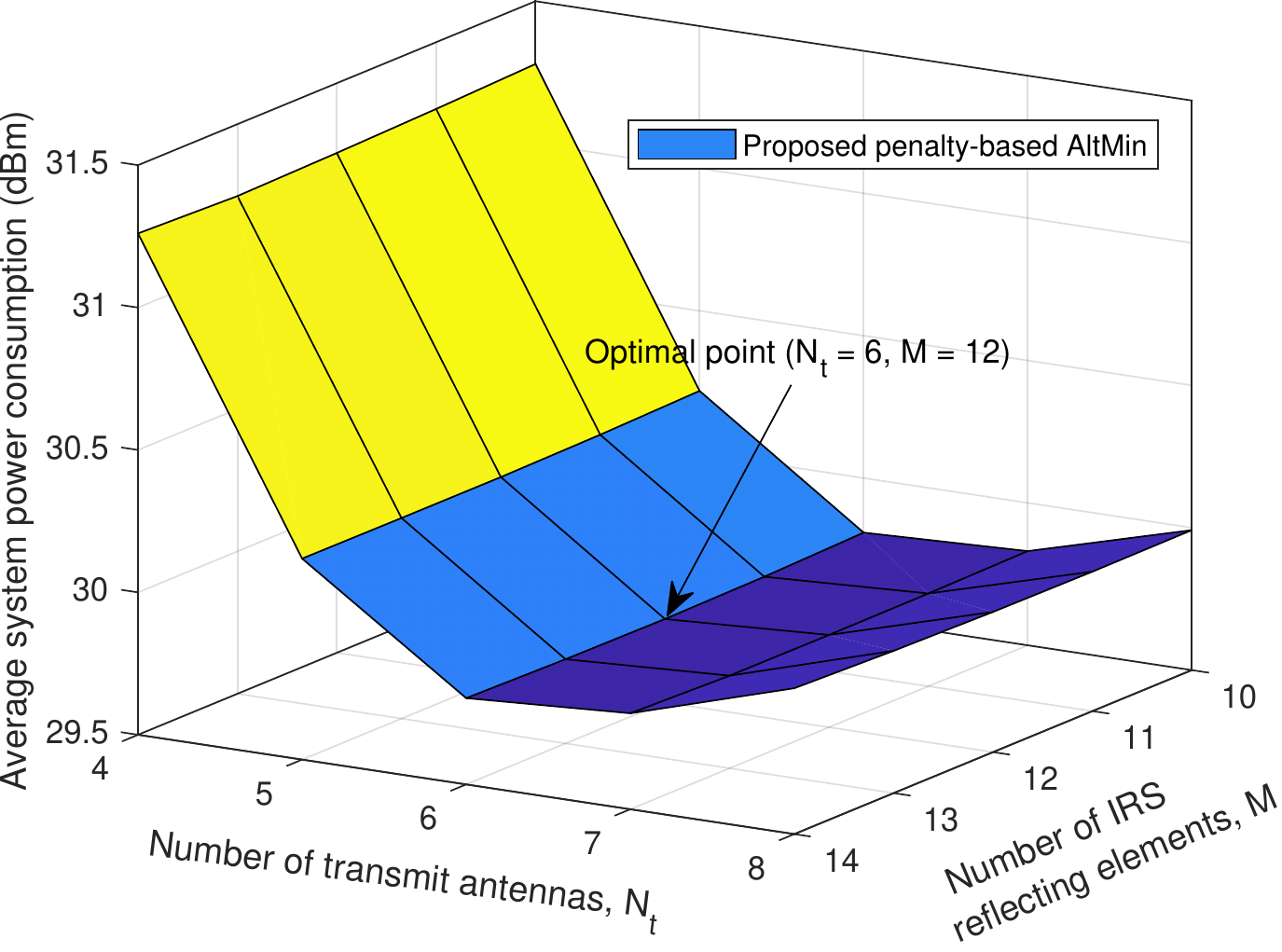}
		\caption{Average system power consumption achieved by the penalty-based AltMin algorithm when $R=80$ m, $K=3$, $\gamma=2$ dB, and $\kappa=0.02$.} \label{fig9}
	\end{minipage}
\end{figure}

In addition to the  transmit power at the AP, we investigate the total power consumption of the considered IRS-assisted system by also taking into account the powers consumed by the RF chains and IRS controllers, respectively. Specifically, the  system   power consumption is defined as
$P_\mathrm{total}=\frac{1}{\eta}P+P_\mathrm{s}+\Nt P_\mathrm{RF}+MP_\mathrm{IRS}$, where $\eta$ is the power amplifier efficiency, $P_\mathrm{RF}$ and $P_\mathrm{IRS}$ are the fixed powers needed to feed one RF antenna and to control one IRS reflecting element, $P$ is the minimized transmit power,  and $P_\mathrm{s}$ is the static
circuit power of the AP. Following \cite{8888223}, in this subsection, we set $P_\mathrm{s}=34$ mW, $P_\mathrm{RF}=80$ mW, and $P_\mathrm{IRS}=5$ mW and we assume $\eta=1$ for simplicity.
In Fig. \ref{fig8}, we assume that perfect CSI is available at the AP and plot the average   system  power consumption when $\Nt+n$ transmit antennas and $M+n$ IRS reflecting elements are deployed, where $n$ is the increment of antenna/IRS elements.
As can be observed from Fig. \ref{fig8}, the two proposed algorithms result in  lower   system  power consumption compared to the three baseline schemes, which again confirms the effectiveness of our proposals and their capabilities of achieving green wireless communications.
Furthermore, the average   system  power consumption first decreases  with  increasing numbers of  transmit antennas and IRS reflecting elements and then increases when $n$ is large. 
In particular, when $n$ starts to increase, the transmit power, $P$,  is significantly reduced due to the availability of more DoFs facilitated by the larger sizes of the antennas arrays and the IRS.
On the other hand, for large values of $n$,  more power-hungry RF chains and IRS controller components are required to drive the additional antennas and IRS reflecting elements, and the related power consumption, $(\Nt+n) P_\mathrm{RF}+(M+n)P_\mathrm{IRS}$, outweighs the reduction in transmit power, $P$, facilitated by deploying more antennas and IRS reflecting elements. 
This leads to an increase in the total power consumption of the IRS-assisted system. Therefore, if  system  power consumption minimization is desired, the numbers of transmit antennas and IRS reflecting elements have to be carefully chosen. We note that the minimum system power consumption can be found by solving the considered problem for different values of $\Nt$ and $M$. Fig. \ref{fig9} depicts the average system power consumption as a function of the numbers of transmit antennas and IRS reflecting elements when the proposed penalty-based AltMin algorithm is applied to combat the CSI uncertainty. As we inferred from Fig. \ref{fig8}, there exists an optimal  pair of $\Nt$ and $M$ that minimizes the average system power consumption, i.e.,  $\Nt=6$ and $M=12$. Since the power consumed by an RF chain is typically several times higher than that for controlling one IRS reflecting element, it is advantageous to deploy more IRS reflecting elements than transmit antennas for minimizing the system power consumption.  
In addition, as can be observed in Fig. \ref{fig9}, the system power consumption $P_\mathrm{total}$ is more sensitive to the number of transmit antennas, $\Nt$, than to the number of IRS reflecting elements, $M$. This observation indicates that a near-optimal pair of $\Nt$ and $M$ can be practically obtained by first searching for the optimal $\Nt$ and then optimizing $M$, which avoids the computationally-expensive grid search required for jointly optimizing the numbers of transmit antennas and IRS reflecting elements.

\section{Conclusions}
In this paper, IRSs were leveraged for realizing green multiuser MISO wireless communication. The total transmit power was minimized while taking into account the QoS requirements of the users. 
Two novel algorithms, i.e., the proposed penalty-based AltMin and IA algorithms, were proposed to jointly optimize the beamforming vectors at the AP and the phase shifts at the IRSs when the CSI is perfectly known at the AP.
Different from existing algorithms that cannot guarantee convergence, one particular contribution of this paper is that the penalty-based AltMin algorithm is guaranteed fast convergence to a stationary point while the IA algorithm ensures a locally optimal KKT solution. 
Furthermore, in the presence of CSI uncertainty, the penalty-based AltMin algorithm was extended to yield a stationary point of the formulated non-convex robust optimization problem. 
Our simulation results verified the significant potential of IRSs to enable green wireless communication. In addition, our simulation results confirmed the effectiveness of the two proposed algorithms for the perfect CSI scenario and, more importantly, the robustness of the proposed penalty-based AltMin algorithm against CSI imperfection.
Finally, system design insights were revealed via simulations. Specifically, it was shown that equipping a small number of transmit antennas and a relatively large number of IRS reflecting elements is beneficial to minimize the total power consumption of  IRS-assisted systems, thereby facilitating green wireless communication.
%
\appendix
\subsection{Proof of Theorem 1}\label{appA}
By relaxing the rank-one constraint \mbox{C4} in problem \eqref{overall}, the remaining problem is jointly convex with respect to the optimization variables and satisfies Slater's constraint qualification.
Hence, strong duality holds and the Lagrangian function  is given by
\begin{equation}
\begin{split}
\mathcal{L}_1&=\sum_{k\in\mathcal{K}}\left[\Tr\left(\mathbf{W}_k\right)+\frac{\delta_k}{2}\left\Vert\tilde{\mathbf{W}}_k+\mathbf{H}_k\mathbf{V}\mathbf{H}_k^H\right\Vert_F^2-\delta_k\Tr\left(\tilde{\mathbf{W}}_k^{(t)}\tilde{\mathbf{W}}_k\right)
-\Tr\left(\mathbf{Y}_k\mathbf{W}_k\right)\right]+\upsilon_1,
\end{split}
\end{equation}
where $\upsilon_1$ comprises all terms that do not involve $\mathbf{W}_k$. $\delta_k\ge0$ and $\mathbf{Y}_k\in\mathbb{H}^\Nt$ are the Lagrange multipliers associated with constraints $\widetilde{\mbox{C1}}$ and $\mathbf{W}_k\succeq\mathbf{0}$, respectively. Then, we reveal the structure of $\mathbf{W}_k$ by examining the relevant KKT conditions of problem \eqref{overall} without \mbox{C4}, which are given by  
\begin{equation}
\mbox{K1:}\,\delta_k^\mathrm{opt}\ge0,\,\mathbf{Y}_k^\mathrm{opt}\succeq\mathbf{0},\quad
\mbox{K2:}\,\mathbf{Y}_k^\mathrm{opt}\mathbf{W}_k^\mathrm{opt}=\mathbf{0},\quad\mbox{K3:}\,\nabla_{\mathbf{W}_k}\mathcal{L}_1\left(\mathbf{W}_k^\mathrm{opt}\right)=\mathbf{0},\quad\forall k.
\end{equation}
With some basic algebraic manipulations,  KKT condition \mbox{K3} can be rewritten as
\begin{equation}\label{eq22}
\mathbf{Y}_k^\mathrm{opt}=\mathbf{I}_\Nt-\boldsymbol{\Delta}_k^\mathrm{opt},
\end{equation}
where 
$\boldsymbol{\Delta}_k^\mathrm{opt}=\delta_k^\mathrm{opt}\left(\tilde{\mathbf{W}}_k^\mathrm{opt}+\mathbf{H}_k\mathbf{V}\mathbf{H}^H_k-\tilde{\mathbf{W}}_k^{(t)}\right)
-\sum_{j\in\mathcal{K}\backslash\{k\}}\delta_j^\mathrm{opt}\gamma_j\left(\tilde{\mathbf{W}}_j^\mathrm{opt}+\mathbf{H}_j\mathbf{V}\mathbf{H}^H_j-\tilde{\mathbf{W}}_j^{(t)}\right)$.
Next, by unveiling the structure of $\mathbf{Y}^\mathrm{opt}_k$, we show that the optimal $\mathbf{W}_k^\mathrm{opt}$ always satisfies $\Rank\left(\mathbf{W}_k^\mathrm{opt}\right)\le1$.
We note that due to the randomness of the channels, the probability of having multiple eigenvalues with the same value $\lambda_{\max}\left(\boldsymbol{\Delta}_k^\mathrm{opt}\right)$ is zero. Reviewing \eqref{eq22}, if $\lambda_{\max}\left(\boldsymbol{\Delta}_k^\mathrm{opt}\right)>1$, then $\mathbf{Y}^\mathrm{opt}_k\succeq\mathbf{0}$ does not hold, which contradicts \mbox{K1}. On the other hand, if $\lambda_{\max}\left(\boldsymbol{\Delta}_k^\mathrm{opt}\right)\le1$, then $\mathbf{Y}^\mathrm{opt}_k$ is a positive semidefinite matrix with $\Rank\left(\mathbf{Y}_k^\mathrm{opt}\right)\ge\Nt-1$, which leads to $\Rank\left(\mathbf{W}_k^\mathrm{opt}\right)\le1$ due to \mbox{K2}. This completes the proof of Theorem 1.


\subsection{Proof of Theorem 2}\label{appB}
By relaxing the rank-one constraint \mbox{C4} in problem \eqref{p332}, the remaining problem is convex with respect to $\mathbf{W}_k$ and satisfies Slater's constraint qualification. Therefore, strong duality holds and the Lagrangian function is given by
\begin{equation}
\mathcal{L}_2=\sum_{k\in\mathcal{K}}\left[\Tr\left(\mathbf{W}_k\right)-\Tr\left(\mathbf{Y}_k\mathbf{W}_k\right)+\Tr\left(\mathbf{Z}_k\mathbf{G}^H\left(\mathbf{V}^T\otimes\tilde{\mathbf{W}}_k\right)\mathbf{G}_k\right)\right]+v_2,
\end{equation}
where $\mathbf{Z}_k$ is the Lagrangian multiplier matrix associated with constraint $\overline{\mbox{C5}}$ and $v_2$ represents the collection of terms that do not depend on $\mathbf{W}_k$. The KKT conditions for problem \eqref{p332} without constraint \mbox{C4} are given by
\begin{equation}
\mbox{T1:}\,\mathbf{Y}_k^\mathrm{opt}\succeq\mathbf{0},\mathbf{Z}_k^\mathrm{opt}\succeq\mathbf{0},\quad\mbox{T2:}\,\mathbf{Y}_k^\mathrm{opt}\mathbf{W}_k^\mathrm{opt}=\mathbf{0},\quad\mbox{T3:}\,\nabla_{\mathbf{W}_k}\mathcal{L}_2\left(\mathbf{W}_k^\mathrm{opt}\right)=\mathbf{0},\quad\forall k.
\end{equation}
 KKT condition \mbox{T3} can be recast as
\begin{equation}\label{eq41}
\mathbf{Y}^\mathrm{opt}_k=\mathbf{I}_{\Nt}-\boldsymbol{\Theta}_k^\mathrm{opt},
\end{equation}
where $\boldsymbol{\Theta}_k^\mathrm{opt}=\mathbf{T}_k^\mathrm{opt}-\sum_{j\in\mathcal{K}\backslash\{k\}}\gamma_j\mathbf{T}_j^\mathrm{opt}$. 
In addition, matrix $\mathbf{T}_k$ is given by
\begin{equation}
\mathbf{T}_k^\mathrm{opt}=\sum_{i=1}^{M+1}\sum_{j=1}^{M+1}v_{ij}\mathbf{U}_{ij}^\mathrm{opt},
\end{equation}
 where $v_{ij}$ is the $(i,j)$-th element of $\mathbf{V}$ and $\mathbf{U}_{ij}^\mathrm{opt}\in\mathbb{C}^{\Nt\times\Nt}$ is the $(i,j)$-th submatrix of $\mathbf{G}_k\mathbf{Z}_k^\mathrm{opt}\mathbf{G}_k^H$, i.e.,\newpage
\begin{equation}
\mathbf{G}_k\mathbf{Z}_k^\mathrm{opt}\mathbf{G}_k^H=
\begin{bmatrix}
\mathbf{U}^\mathrm{opt}_{11}&\mathbf{U}^\mathrm{opt}_{12}&\cdots&\mathbf{U}^\mathrm{opt}_{1(M+1)}\\
\mathbf{U}^\mathrm{opt}_{21}&&&\\
\vdots&&\ddots&\\
\mathbf{U}^\mathrm{opt}_{(M+1)1}&&&\mathbf{U}^\mathrm{opt}_{(M+1)(M+1)}
\end{bmatrix}.
\end{equation}
By applying a similar analysis as in Appendix \ref{appA} to \eqref{eq41}, it can be shown that $\Rank\left(\mathbf{W}_k\right)\le1$ can always be obtained, which completes the proof.

\bibliographystyle{IEEEtran}
\bibliography{jrnl}
\end{document}